\documentclass[hidelinks,journal]{IEEEtran}
\IEEEoverridecommandlockouts
\ifCLASSINFOpdf
\else
\usepackage[dvips]{graphicx}
\fi
\usepackage{url}

\usepackage[dvipsnames]{xcolor}
\usepackage{graphicx} 
\usepackage{booktabs} 
\usepackage{mathrsfs}
\usepackage{amsmath,amssymb,amsfonts}
\usepackage{textcomp}
\usepackage{outlines}

\usepackage{mathtools}
\usepackage{cite}
\usepackage[acronym,shortcuts]{glossaries}
\usepackage{enumitem} 
\setlist[itemize]{noitemsep} 
\usepackage{bm}
\usepackage{comment}
\usepackage{algorithm, algpseudocode}
\usepackage{mathtools}
\usepackage{lipsum}
\usepackage{mleftright}
\usepackage{orcidlink}
\usepackage{cancel}
\usepackage{scalefnt}
\usepackage{relsize}
\usepackage{soul}
\usepackage[bottom]{footmisc}
\usepackage{dblfloatfix}
\usepackage[flushleft]{threeparttable} 

\DeclareMathOperator*{\argmin}{arg\,min}
\def\thesubsubsectiondis{\unskip\arabic{subsubsection})}

\providecommand{\UL}{\mathrel{\raise-4pt\hbox{\hglue -2.8ex
\vrule height .1ex width 2.3ex
\vrule height 3ex width .1ex
\hglue .4ex}}}
\providecommand{\ul}{\mathrel{\raise-2pt\hbox{\hglue -2.3ex
\vrule height .1ex width 2ex
\vrule height 2ex width .1ex
\hglue .4ex}}}
\providecommand{\ub}{
\mathrel{\hbox{\hglue -2.8ex \vrule height 2ex width .06ex}
\raise2ex\hbox{\hglue -0.1ex\vrule height .1ex width 2.5ex}
\hbox{\hglue -0.1ex \vrule height 2ex width .1ex
\hglue .4ex}}}
\providecommand{\lb}{
\mathrel{\raise-0.5ex
\hbox{\hglue -2.8ex \vrule height 2ex width .1ex
\vrule height .1ex width 2.5ex
\vrule height 2ex width .1ex
\hglue .4ex}}}
\providecommand{\UB}{
\mathrel{
\raise-1ex\hbox{\hglue -1.8em \vrule height 3ex width .1ex}
\raise2ex\hbox{\hglue -0.1ex\vrule height .1ex width 1.5em}
\raise-1ex\hbox{\hglue -0.1ex \vrule height 3ex width .1ex
\hglue .4ex}}}
\providecommand{\LB}{
\mathrel{\raise-1ex
\hbox{\hglue -1.8em \vrule height 3ex width .1ex
\vrule height .1ex width 1.5em
\vrule height 3ex width .1ex
\hglue .4ex}}}
\usepackage{trimclip}
\newif\iflclip
\newif\ifbclip
\newif\ifrclip
\newif\iftclip
\def\CLIP{\dimexpr\fboxrule+.2pt\relax}
\def\nulclip{0pt}
\newcommand\partbox[2]{%
\lclipfalse\bclipfalse\rclipfalse\tclipfalse%
\let\lkern\relax\let\rkern\relax%
\let\lclip\nulclip\let\bclip\nulclip\let\rclip\nulclip\let\tclip\nulclip%
\parseclip#1\relax\relax%
\iflclip\def\lkern{\kern\CLIP}\def\lclip{\CLIP}\fi
\ifbclip\def\bclip{\CLIP}\fi
\ifrclip\def\rkern{\kern\CLIP}\def\rclip{\CLIP}\fi
\iftclip\def\tclip{\CLIP}\fi
\lkern\clipbox{\lclip{} \bclip{} \rclip{} \tclip}{\fbox{#2}}\rkern%
}
\def\parseclip#1#2\relax{%
\ifx l#1\lcliptrue\else
\ifx b#1\bcliptrue\else
\ifx r#1\rcliptrue\else
\ifx t#1\tcliptrue\else
\fi\fi\fi\fi
\ifx\relax#2\relax\else\parseclip#2\relax\fi
}
\usepackage{efbox}

\newcommand\Tstrut{\rule{0pt}{2.6ex}}         

\newacronym{IoV}{IoV}{internet of vehicles}
\newacronym{SVD}{SVD}{singular value decomposition}
\newacronym{DCM}{DCM}{double-centering matrix}
\newacronym{SNR}{SNR}{signal-to-noise ratio}
\newacronym{3D}{3D}{three-dimensional}
\newacronym{GA}{GA}{genie-aided}
\newacronym{EA}{EA}{``\emph{estimate-then-average}''}
\newacronym{AE}{AE}{``\emph{average-then-estimate}''}
\newacronym{IRS}{IRS}{intelligent reflecting surface}
\newacronym{RSSI}{RSSI}{received signal strength indicator}
\newacronym{SotA}{SotA}{state-of-the-art}
\newacronym{CSI}{CSI}{channel state information}
\newacronym{D2D}{D2D}{device-to-device}
\newacronym{RR}{RR}{round-robin}
\newacronym{DA}{DA}{Dutch auction}
\newacronym{AV}{AV}{autonomous vehicle}
\newacronym{CWFL}{CWFL}{clustered WFL}
\newacronym{WFL}{WFL}{wireless federated learning}
\newacronym{RSMA}{RSMA}{rate splitting multiple access}
\newacronym{IoT}{IoT}{Internet-of-Things}
\newacronym{TDMA}{TDMA}{time-domain multiple access}
\newacronym{NOMA}{NOMA}{non-orthogonal multiple access}
\newacronym{ML}{ML}{machine learning}
\newacronym{MIMO}{MIMO}{multiple-input multiple-output}
\newacronym{CT}{CT}{compute-then-transmit}
\newacronym{FP}{FP}{fractional programming}
\newacronym{CF-mMIMO}{CF-mMIMO}{cell free massive MIMO}
\newacronym{iid}{i.i.d.}{independent and identically distributed}
\newacronym{AD}{AD}{autonomous driving}
\newacronym{DL}{DL}{downlink}
\newacronym{UL}{UL}{uplink}
\newacronym{IC}{IC}{interference cancellation}
\newacronym{SIC}{SIC}{successive interference cancellation}
\newacronym{BS}{BS}{base station}
\newacronym{TX}{TX}{transmit}
\newacronym{RX}{RX}{receive}
\newacronym{MU}{MU}{multi-user}
\newacronym{SISO}{SISO}{single-input single-output}
\newacronym{AWGN}{AWGN}{additive white Gaussian noise}
\newacronym{SINR}{SINR}{signal-to-interference-and-noise ratio}
\newacronym{FL}{FL}{federated learning}
\newacronym{CPU}{CPU}{central processing unit}
\newacronym{KNN}{KNN}{K-nearest-neighbor}
\newacronym{RF}{RF}{radio frequency}
\newacronym{GD}{GD}{gradient descent}
\newacronym{V2X}{V2X}{vehicle-to-anything}
\newacronym{DT}{DT}{digital twins}

\newacronym{RSS}{RSS}{received signal strength}
\newacronym{FIM}{FIM}{fisher information matrix}
\newacronym{ToA}{ToA}{time of arrival}
\newacronym{ToF}{ToF}{time of flightl}
\newacronym{AoA}{AoA}{angle of arrival}
\newacronym{GP}{GP}{Gaussian process}
\newacronym{2D}{2D}{two-dimensional}
\newacronym{GPR}{GPR}{Gaussian process regression}
\newacronym{GNSS}{GNSS}{global navigation satellite systems}
\newacronym{B5G}{B5G}{beyond fifth-generation}
\newacronym{6G}{6G}{sixth-generation}
\newacronym{RRH}{RRH}{remote radio head}
\newacronym{GPS}{GPS}{Global Positioning System}
\newacronym{RFID}{RFID}{radio frequency identification}
\newacronym{TCAS}{TCAS}{traffic alert and collision avoidance systems}
\newacronym{RMSE}{RMSE}{root mean square error}
\newacronym{SGD}{SGD}{stochastic gradient descent}
\newacronym{PDF}{PDF}{probability density function}
\newacronym{CU}{CU}{computing unit}
\newacronym{DM-MIMO}{DM-MIMO}{distributed massive multiple-input multiple-output}
\newacronym{LOS}{LOS}{line-of-sight}
\newacronym{NLOS}{NLOS}{non-line-of-sight}
\newacronym{ROI}{ROI}{region of interest}
\newacronym{AP}{AP}{access point}
\newacronym{TDOA}{TDOA}{time difference of arrival}
\newacronym{UE}{UE}{user equipment}
\newacronym{dB}{dB}{decibel}
\newacronym{RIS}{RIS}{reconfigurable intelligent surface}
\newacronym{CG}{CG}{conjugate gradient}

\newacronym{PG}{PG}{proximal gradient}
\newacronym{SVT}{SVT}{singular value thresholding}
\newacronym{NN}{NN}{nuclear norm}
\newacronym{NMSE}{NMSE}{normalized mean square error}
\newacronym{MC}{MC}{matrix completion}
\newacronym{NP}{NP}{non-deterministic polynomial-time}
\newacronym{EDM}{EDM}{euclidean distance matrix}
\newacronym{SC}{SC}{soft-connected}
\newacronym{CRLB}{CRLB}{Cramér-Rao Lower Bound}
\newacronym{PoA}{PoA}{phase of arrival}
\newacronym{UAV}{UAV}{unmanned aerial vehicle}
\newacronym{VR}{VR}{virtual reality}
\newacronym{MDS}{MDS}{multidimensional scaling}

\newacronym{RBL}{RBL}{rigid body localization}
\newacronym{RBT}{RBT}{rigid body tracking}
\newacronym{SC-RBL}{SC-RBL}{soft-connected RBL}
\newacronym{W-RBL}{W-RBL}{\underline{wireless} RBL}

\newacronym{SDP}{SDP}{semidefinite programming}
\newacronym{JCAS}{JCAS}{joint communication and sensing}
\newacronym{SDR}{SDR}{semi-definite relaxation}

\newacronym{OPP}{OPP}{orthogonal Procrustes problem}
\newacronym{SLAM}{SLAM}{simultaneous localization and mapping}
\newacronym{WLS}{WLS}{weighted least square}
\newacronym{SI}{SI}{soft-impute}

\begin{document}

\title{ $~$ \\[-3ex] {\normalsize{\textbf{\color{red}Please find the official IEEE \underline{published} version of this article on IEEE Xplore \href{https://ieeexplore.ieee.org/document/11270017}{{\color{blue}[here]}}} \textbf{\color{red}and cite as:} \\} 

\color{cyan} N F\"uhrling \emph{et al.}, ``Robust Egoistic Rigid Body Localization,"\\[-4ex]

\emph{in IEEE Transactions on Signal Processing}, vol. 73, pp. 5076-5089, Nov. 2025}\\[-0.5ex]

Robust Egoistic Rigid Body Localization\vspace{-.5ex}}

\author{Niclas~F\"uhrling\textsuperscript{\orcidlink{0000-0003-1942-8691}}, \IEEEmembership{Graduate Student Member, IEEE}, Giuseppe~Thadeu~Freitas~de~Abreu\textsuperscript{\orcidlink{0000-0002-5018-8174}}, \IEEEmembership{Senior Member, IEEE}\\ David~Gonz{\'a}lez~G.\textsuperscript{\orcidlink{0000-0003-2090-8481}}, \IEEEmembership{Senior Member, IEEE} and Osvaldo~Gonsa\textsuperscript{\orcidlink{0000-0001-5452-8159}}\vspace{-3ex}

\thanks{N.~F\"uhrling and G.~T.~F.~Abreu are with the School of Computer Science and Engineering, Constructor University, Campus Ring 1, 28759, Bremen, Germany (emails: [nfuehrling, gabreu]@constructor.university).}
\thanks{D.~Gonz{\'a}lez~G. and O.~Gonsa are with the Wireless Communications Technologies Group, Continental AG, Wilhelm-Fay Strasse 30, 65936, Frankfurt am Main, Germany (e-mails: david.gonzalez.g@ieee.org, osvaldo.gonsa@continental-corporation.com).}
\thanks{Parts of this article have been {\color{black}published at} the 2025 IEEE Wireless Communications and Networking Conference (WCNC) \cite{Nic_RBL} (Corresponding author: N. F\"uhrling)}
\vspace{-2ex}
}

\markboth{ }
{Shell \MakeLowercase{\textit{et al.}}: Bare Demo of IEEEtran.cls for IEEE Journals}
\maketitle

\begin{abstract}

We consider a robust and self-reliant (or ``egoistic'') variation of the \ac{RBL} problem, in which a primary rigid body seeks to estimate the pose ($i.e.$, location and orientation) of another rigid body (or ``target''), relative to its own, without the assistance of external infrastructure, without prior knowledge of the shape of the target, and taking into account the possibility that the available observations are incomplete.
Three complementary contributions are then offered for such a scenario. 
The first is a method to estimate the translation vector between the center {\color{black}points} of both rigid bodies, which unlike existing techniques does not require that both objects have the same shape or even the same number of landmark points.
This technique is shown to significantly outperform the \ac{SotA} under complete information, but to be sensitive to data erasures, even when enhanced by matrix completion methods.
The second contribution, designed to offer improved performance in the presence of incomplete information, offers a robust alternative to the latter, at the expense of a slight relative loss under complete information.
Finally, the third contribution is a scheme for the estimation of the rotation matrix describing the relative orientation of the target rigid body with respect to the primary.
Comparisons of the proposed schemes and \ac{SotA} techniques demonstrate the advantage of the contributed methods in terms of \ac{RMSE} performance under fully complete information and incomplete conditions.
\end{abstract}


\IEEEpeerreviewmaketitle

\vspace{-2ex}
\section{Introduction}

\IEEEPARstart{W}{ireless} localization \cite{burghal_2020, obeidat2021review, ShanAccess2023} can be seen as a precursor of \ac{JCAS}, in so far as it demonstrates that communication signals can be used to locate users and acquire situation awareness, functionalities that have been identified as key drivers for \ac{B5G} \cite{WangJSAC2022} and \ac{6G} \cite{02:00074} systems, as well as new applications such as the \ac{IoV} \cite{WangTIV2020} and digital twins \cite{ManickamAcces2023}.

Indeed, there are several types of information -- including finger-prints \cite{VoCST2016}, \ac{RSSI} \cite{Nic:RSSI}, \ac{AoA} \cite{Al-SadoonTAP2020}, and delay-based estimates of radio range \cite{ZengTSP2022} -- can be extracted from radio signals for the purpose of localization.
In much of related literature it is considered that acquiring such information  requires specialized equipment, dedicated protocols, and/or the transmission of purpose-designed signals, which explains the predominance of methods to find the position of individual points \cite{Yassin_2016, obeidat2021review, burghal_2020}, given the additional overhead, costs and other constraints.

Recently, however, advances in \ac{JCAS} technology \cite{Zhang_2021} {\color{black}have} demonstrated that radar parameters ($i.e.$, range, bearing and velocity) can be acquired by conventional communications signals \cite{Rayan_2024,Rayan_Journal}, not only actively, $i.e.$, using signals transmitted by the target to the sensors, but also passively, $i.e.$, using round-trip reflections of signals transmitted by the sensors themselves, which in turn implies a more abundant and richer availability of positioning information.
A consequence of this recent development is an increasing interest in the \acf{RBL} problem \cite{WangTSP2020,Nic_RBL_WP,FuehrlingV2X2024}, whose objective is to determine not only the location of point targets, or their average, but the shape and orientation of objects, based on a collection of points sufficient to define the latter.

\Acl{RBL} is particularly attractive to \ac{V2X} networks and \ac{IoV} applications which, unlike earlier applications of positioning technology such as people tracking in indoor settings \cite{Yassin_2016} and asset management in industrial settings \cite{AHMED_2020}, crucially require information on the size, shape, and orientation of vehicles in order to ensure the efficacy and safety of \acf{AD} applications such as collision detection \cite{Bruk_2023}, navigation \cite{eckenhoff_2019}, and vehicle path prediction \cite{Huang_2022}, to name only a few examples.

Focusing on the \ac{V2X} and \ac{IoV} paradigm in particular, a scenario commonly encountered is such that a vehicle is able to obtain relative information between itself and surrounding vehicles, which if processed adequately can be utilized to enable \ac{RBL} as a means to enrich applications such as advanced \ac{AD}, platooning and more.

We emphasize that for such a purpose conventional techniques, such as \ac{SLAM} technologies \cite{Huang_2019, Barros_2022, Bavle_2023} are not suitable, as they require expensive dedicated equipment and massive amounts of data to function, implicating in high costs and latencies that limit their feasibility.
To extend such techniques further, complementary sensing modalities, such as LiDAR or vision-based systems \cite{Yin2024,chen2021lidarlocalization,Elhousni_2020} can be integrated. 
These sensors provide rich geometric information that can refine or validate the range-based pose estimates, but in conditions, such as dust, fog, or occlusions the performance may degrade. 
Such hybrid sensing strategies could enable robust fusion multi-sensor localization for autonomous systems operating in GPS-denied or visually degraded environments.
However, in contrast to \ac{SLAM} or LiDAR, the type of \ac{RBL} problem here addressed is based on radio signals, preferably under (but not limited to) a \ac{JCAS} paradigm {\color{black}\cite{Chen_2015,9447218,9904904}}.
Examples of the latter are the method proposed in \cite{Bras_2016}, where the pose, angular velocity and trajectory of a rigid body is estimated using Lyapunov functions of Doppler measurements, obtained by a nonlinear observer; the technique in \cite{Chen_2015}, where a two-stage algorithm was used to estimate rotation, translation, angular velocity and translational velocity from range and Doppler measurements, making use of various \ac{WLS} minimization methods; and the scheme in \cite{PizzoICASSP2016}, which proposes a solution to the relative multi-object \ac{RBL} problem\footnote{An anchor-based version of the method also appeared earlier in \cite{Chepuri_2013}.} in an anchorless scenario, whereby the relative translation and rotation between two rigid bodies is estimated by measuring the cross-body \ac{LOS} distances between the points defining the two bodies.

\vspace{-3ex}
{
\subsection{Scope and Contributions}
\vspace{-1ex}

While previous studies on rigid body localization have addressed translation and rotation estimation under ideal conditions, they generally rely on prior knowledge of the target shape \cite{Chepuri_2013, PizzoICASSP2016,9186663}, which is unrealistic in real life applications since vehicles vary greatly in shape and size.
In addition, a recurrent problem in localization systems which is even more critical in the \ac{RBL} scenario, is that measurements are often missing due to channel blockage, poor \ac{SNR}, packet losses, etc., which significantly limits their applicability in practical vehicular and \ac{IoV} scenarios.

In light of all the above, this work addresses these gaps by focusing on the estimation of relative pose between two rigid bodies, $i.e.$, translation and rotation parameters, via a robust and egoistic radio-based \ac{RBL} schemes using only range-based measurements, without assuming knowledge of the target's geometry or requiring external infrastructure.
Additionally, the proposed robust method is capable of operating with incomplete information, which offers a reliable, low-cost alternative to \ac{SLAM} {or LiDAR}.
The contributions of the article aimed at closing this research gap can be summarized as follows:
\begin{itemize}
\item A novel rigid body localization scheme for the estimation of the translation vector between two rigid bodies of arbitrary shapes is described. In contrast to existing solutions \cite{Chen_2015,9447218,9904904,Bras_2016, Chepuri_2013, PizzoICASSP2016,9186663}, the proposed method enables the first body to act in a self-reliant (``egoistic'') manner, in the sense that no knowledge of the shape of the second (``target'') body is needed by the primary rigid body to perform the estimation.
The proposed technique is found to outperform the most closely related \ac{SotA} alternative, even in scenarios with incomplete observations.
\item A second new translation vector estimation method is proposed, which extends\footnote{Besides the extension, we also correct an error made in \cite[Subsec. 3.2]{PizzoICASSP2016}, which renders that approach ineffective for the estimation of the translation vector $\boldsymbol{t}$. For details see Subsection \ref{sec:Q_est} and Appendix A.} the technique in \cite{PizzoICASSP2016} from a translation distance estimator to a full translation vector estimator, which is robust to incomplete observations and is fully operational in an egoistic manner, outperforming the first proposed method, as well as the \ac{SotA} \cite{Chen_2015} under such conditions.
\item A scheme for the estimation of the rotation matrix between two rigid bodies is proposed that works in an egoistic manner, and is independent of translation vector estimates, which is used to estimate the full pose of the target rigid body in combination with either of the two proposed translation vector estimation methods.
 
\end{itemize}

Altogether, we propose two complementary egoistic methods for the estimation of the translation vector between two rigid bodies of arbitrary shapes, which can be selected according to the availability of observations, as well as an egoistic rotation matrix estimation method that is independent of the translation vector estimates.

The structure of the remainder of the article is as follows.
First, Section \ref{sec:prior} describes the system model, concisely and clearly stating the problem mathematically, and briefly elaborating on limitations of related \ac{SotA} approaches.
The two proposed egoistic translation vector estimation methods are described and evaluated in Sections \ref{sec:prop} and \ref{sec:prop_journal}, respectively, while the egoistic rotation estimation method is described in Section \ref{sec:Q_C_est}, along with a brief complexity analysis.
}

\vspace{-2ex}
\section{Rigid Body Localization: System Model, Problem Formulation and \Acl{SotA}}
\label{sec:prior}
\vspace{-1ex}

\subsection{System Model}

\begin{figure}[H]
\centering
\includegraphics[width=\columnwidth]{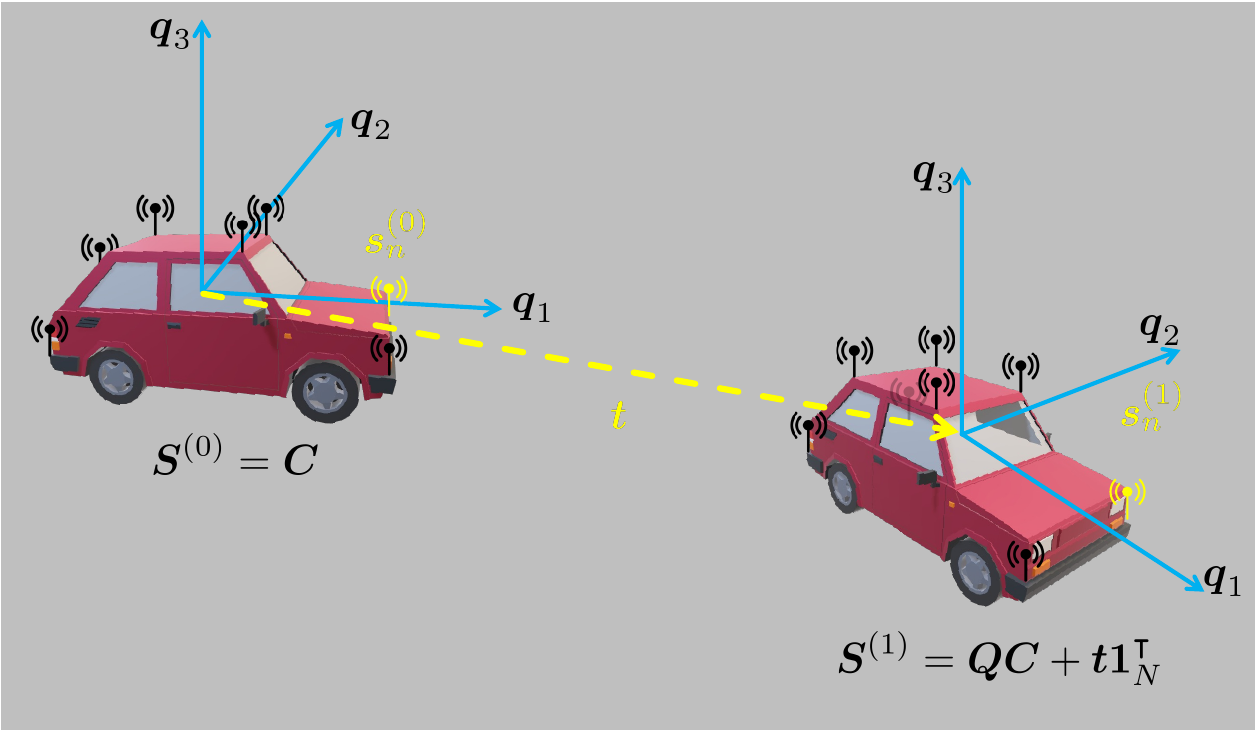}
\vspace{-4ex}
\caption{Illustration of a rigid body transformation at two distinct locations $\boldsymbol{S}^{(0)}$ and $\boldsymbol{S}^{(1)}$. Without loss of generality, we set the initial position to be identical to the matrix $\boldsymbol{C}$, which defines the shape and orientation of the rigid body. The second location $\boldsymbol{S}^{(1)}$ of the body is then determined according to equation \eqref{eq:basic_model_one_body}, and is obtained by the transformation of $\boldsymbol{S}^{(0)}$ via a rotation matrix $\bm{Q}$ and a translation vector $\boldsymbol{t}$.}
\label{fig:RB_tra}
\vspace{-2ex}
\end{figure}

Consider a rigid body represented by a collection of $N$ landmark points $\boldsymbol{c}_n\in\mathbb{R}^{3\times 1}$ in the \ac{3D} space, with $n=\{1,\cdots,N\}$, such that the shape of said body is described by the corresponding conformation matrix $\boldsymbol{C}$ constructed by the column-wise collection of the vectors $\boldsymbol{c}_n$, such that $\boldsymbol{C}=[\boldsymbol{c}_{1},\cdots,\boldsymbol{c}_{N}]\in \mathbb{R}^{3\times N}$.
Next, as illustrated in Figure \ref{fig:RB_tra}, consider the representation of the location $\boldsymbol{S}^{(1)}$ of said rigid body relative to another ($e.g.$, previous) location $\boldsymbol{S}^{(0)}$, which without loss of generality can be set as a ``canonical'' reference (centered at the absolute origin),
such that $\boldsymbol{S}^{(0)}=\boldsymbol{C}$.
Then, one can write\footnote{For simplicity, we hereafter omit super-scrips $^{(\cdot)}$ whenever clarity is not compromised.}, without loss of generality
\vspace{-0.5ex}
\begin{equation}
\label{eq:basic_model_one_body}
\boldsymbol{S}=\boldsymbol{Q}\cdot\boldsymbol{C}+\boldsymbol{t}\cdot\boldsymbol{1}_{N}^{\intercal}=\mleft[\boldsymbol{Q}|\boldsymbol{t}\mright]\mleft[
\begin{array}{c}
\bm{C} \\
\hline
\boldsymbol{1}_{N}^{\intercal}
\end{array}
\mright],
\vspace{-0.5ex}
\end{equation}
where $\boldsymbol{t}\in\mathbb{R}^{3\times 1}$ is a translation vector given by the distance between the geometric centers of the body at the two locations, $\boldsymbol{1}_{N}$ is a column vector with $N$ entries all equal to $1$, and $\boldsymbol{Q}\in \mathbb{R}^{3\times 3}$ is a rotation matrix\footnote{For the sake of simplicity, in coherence with the \ac{SotA}, detecting the orientation of a rigid body in this article will be interpreted as estimating of the $9$ elements of the corresponding rotation matrix $\bm{Q}$ as a whole. However, as shown in \cite{vizitivWCNC2025} this representation can be extended by replacing the estimation of $\bm{Q}$ with the estimation of the associated and fundamental yaw, pitch and roll angles $(\alpha, \beta, \gamma)$.} determined by the corresponding yaw, pitch and roll angles $\alpha, \beta$ and $\gamma$, respectively, namely

\vspace{-2ex}
\begin{figure}[H]
\includegraphics[width=\columnwidth]{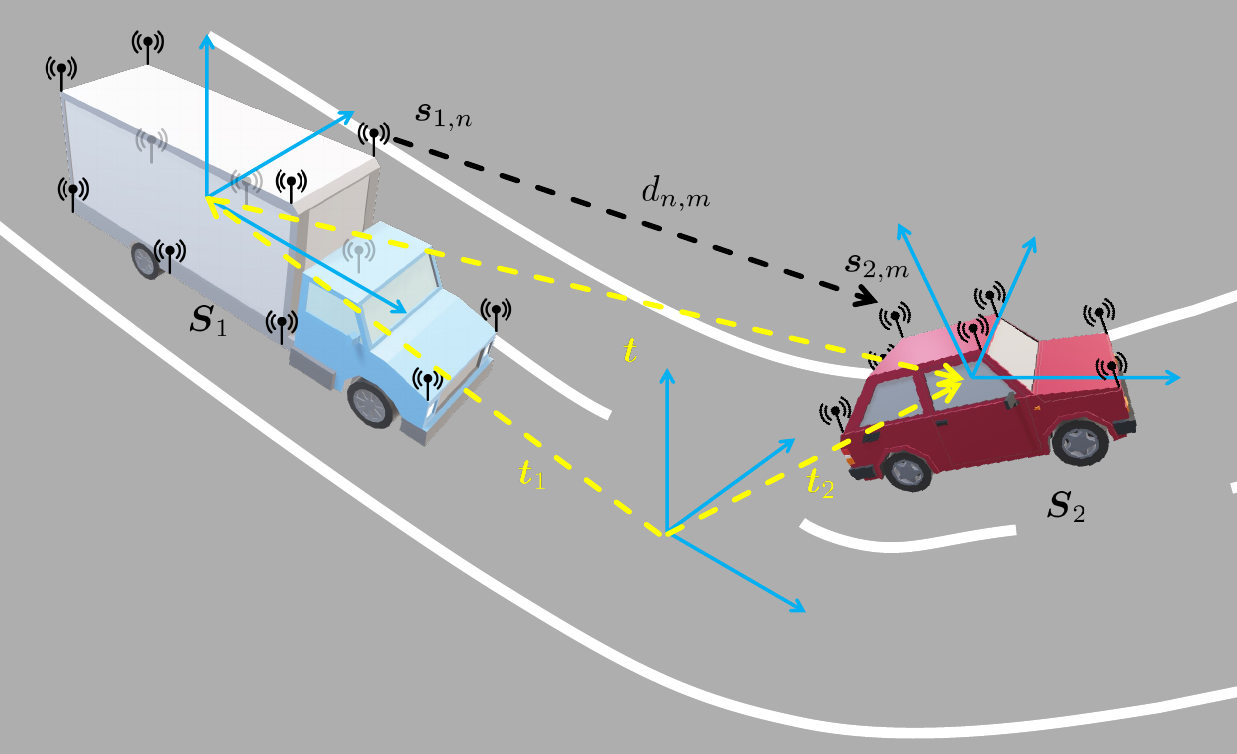}
\vspace{-4ex}
\caption{Illustration of a two-body egoistic \ac{RBL} scenario. Each rigid body has a different shape and orientation, defined by distinct conformation matrices $\boldsymbol{C}_1$ and $\boldsymbol{C}_2$, respectively. The translation vector $\boldsymbol{t}$ between the bodies, highlighted in yellow, is defined by the difference between the geometric centers of the two rigid bodies.}
\label{fig:sys}
\vspace{-4ex}
\end{figure}

\vspace{-0.5ex}
\begin{eqnarray}
\bm{Q} \triangleq \bm{Q}_{z}(\gamma)\,\bm{Q}_{y}(\beta)\,\bm{Q}_{x}(\alpha)&&\\
&&\hspace{-28ex}
=\!\!\!\left[
\begin{array}{@{}c@{\;\,}c@{\;\,}c@{}}
\cos\gamma&-\sin\gamma& 0\\
\sin\gamma& \cos\gamma& 0\\
0 	    & 0           & 1\\
\end{array}\right]\!\!\!\cdot\!\!\!
\left[
\begin{array}{@{}c@{\;\,}c@{\;\,}c@{}}
\cos\beta & 0           & \sin\beta\\
0			& 1			  & 0\\
-\sin\beta& 0 		  & \cos\beta\\
\end{array}\right]\!\!\!\cdot\!\!\!
\left[
\begin{array}{@{\,}c@{\;\,}c@{\;\,}c@{\!}}
1 			& 0			  & 0\\
0			& \cos\alpha& -\sin\alpha\\
0			& \sin\alpha& \cos\alpha\\
\end{array}\right]\nonumber\\
&&\hspace{-28ex}
=\!\!\text{\scalebox{0.75}{$
\left[
\begin{array}{@{\,}c@{\;\,}c@{\;\,}c@{\,}}
\cos\beta\cos\gamma & \sin\alpha\sin\beta\cos\gamma- \cos\alpha\sin\gamma & \cos\alpha\sin\beta\cos\gamma+\sin\alpha\sin\gamma\\
\cos\beta\sin\gamma & \sin\alpha\sin\beta\sin\gamma+ \cos\alpha\cos\gamma & \cos\alpha\sin\beta\sin\gamma-\sin\alpha\cos\gamma\\
-\sin\beta			& \sin\alpha\cos\beta								  & \cos\alpha\cos\beta\\
\end{array}\right]$}}\nonumber\\
&&\hspace{-28ex}
=\!\!\left[
\begin{array}{@{\,}c@{\;\,}c@{\;\,}c@{\,}}
q_{1,1} & q_{1,2} & q_{1,3}\\
q_{2,1} & q_{2,2} & q_{2,3}\\
q_{3,1} & q_{3,2} & q_{3,3}\\
\end{array}\right]\!\!.\nonumber
\end{eqnarray}

Next, consider a scenario as illustrated in Figure \ref{fig:sys}, in which two distinct rigid bodies, hereafter referred to by their indices $i=\{1,2\}$, have generally different shapes and/or are characterized by generally distinct numbers $N_1$ and $N_2$ of landmark points, respectively, such that under a common absolute reference, the bodies are represented by the corresponding distinct conformation matrices $\bm{C}_1\in\mathbb{R}^{3\times N_1}$ and $\bm{C}_2\in\mathbb{R}^{3\times N_2}$.

Since $\bm{C}_1 \neq \bm{C}_2$, it is obvious that in such a scenario the location of one body relative to the other cannot be described in terms of equation \eqref{eq:basic_model_one_body}.
A common scenario in \ac{V2X} systems with relevance to \ac{AD} applications is, however, that one of the rigid bodies -- say, the truck in Figure \ref{fig:sys} -- seeks to estimate not only its distance to the other body -- in this case, the car in Figure \ref{fig:sys} -- but also the shape and orientation of the latter, based on a set of measurements of the distances between their corresponding landmark points.

It will be considered, in what follows, that the distance measurements among the landmark points, hereafter referred to as ``sensors'', on both rigid bodies can be obtained either via point-to-point communications or other technologies such as radar, video or \ac{JCAS}.
It will, furthermore, be assumed that each body is only aware of its own shape, described by the corresponding conformation matrices $\boldsymbol{C}_i=[\boldsymbol{c}_{i,1},\cdots,\boldsymbol{c}_{i,N_i}]\in \mathbb{R}^{3\times N_i}$, where $\boldsymbol{c}_{i,n}$, is the location of the $n$-th point of the $i$-th body, with respect to its geometric center.

When subjected to unbiased estimation errors, the estimates of the pairwise distance between sensors $\bm{s}_{1,n}$ on the first body, and $\bm{s}_{2,m}$ on the second\footnote{For simplicity, we slightly abuse the notation by using $\bm{s}_{i,n}$ to refer both to a sensor and its location.}, can be described by
\vspace{-0.5ex}
\begin{equation}
\label{eq:distance_measurements}
\tilde{d}_{n,m} = d_{n,m} + \upsilon_{n,m},
\vspace{-0.5ex}
\end{equation}
where $d_{n,m} \triangleq ||\boldsymbol{s}_{1,n}-\boldsymbol{s}_{2,m}||_2$ is the true pairwise distance between the sensors, while $\upsilon_{n,m}$ denotes measurement noise modeled as i.i.d. zero mean Gaussian random variables with variance $\sigma^2$.

In order to avoid negative numbers and linearize the relationship between the acquired squared distances and corresponding measurement errors, we shall also consider the equivalent model, given by
\vspace{-0.5ex}
\begin{equation}
\label{eq:sqr_meas}
\tilde{d}^2_{n,m} = d^2_{n,m} + \omega_{n,m},
\vspace{-0.5ex}
\end{equation}
where the mean and variance of the squared-distance estimation error $\omega_{n,m}$ are respectively given by $\mathbb{E}[\omega_{n,m}]=\sigma^2$ and $\mathbb{E}\big[\big(\omega_{n,m}-\mathbb{E}[\omega_{n,m}]\big)^2\big]=4d^2_{n,m}\sigma^2+2\sigma^4$, as described in \cite{Chepuri_2013}.

It proves convenient, to collect the true distances $d_{n,m}$ from above into the \ac{EDM}
\begin{equation}
\color{black}
\bm{D}=\mleft[
\begin{array}{c|c}
\bm{D}_1 & \bm{D}_{12} \\
\hline
\bm{D}_{12}^{\intercal}&\bm{D}_2
\end{array}
\mright] \in\mathbb{R}^{(N_1+N_2)\times (N_1+N_2)},
\label{eq:full_D}
\end{equation}
which includes both the pairwise distances between the rigid bodies $\bm{D}_{12}$, as well as the intra-distances of the two individual rigid bodies $\bm{D}_1$ and $\bm{D}_2$.

Without loss of generality, it can be assumed under the system model described above, that the self intra-distance matrix $\bm{D}_1$ is known exactly, the target intra-distance matrix $\bm{D}_2$ is unknown, and the squared cross-distance matrix $\bm{D}_{12}$ can be written as \cite{Torgerson_1952}

\quad\\[-5ex]
\begin{equation}
\bm{D}_{12}^{\odot 2}=\bm{D}_{12}\odot \bm{D}_{12}=\boldsymbol{\psi}_1\boldsymbol{1}_{N_2}^{\intercal}+\boldsymbol{1}_{N_1}\boldsymbol{\psi}_2^{\intercal}-2\boldsymbol{S}_{1}^{\intercal}\boldsymbol{S}_{2},
\label{eq:meas_full}
\end{equation}
where $\boldsymbol{S}_{1}$ and $\boldsymbol{S}_{2}$ are matrices containing the locations of the sensors in bodies 1 and 2, respectively; the auxiliary vectors $\boldsymbol{\psi}_i \triangleq \boldsymbol{S}_{i}^{\intercal}\boldsymbol{S}_{i}$ carry the squared norms of the corresponding individual sensor locations, and the symbol $\odot$ indicates an element-wise matrix operation ($e.g.$, multiplication or exponentiation).

Since the cross-body measurements typically can only be assured to be available under \ac{LOS} conditions, it is possible that some distances between sensors on both bodies cannot be measured.
To incorporate this into the given model the modified notation of the squared measurements is given by
\begin{equation}
\boldsymbol{D}_{12}^{\odot 2}\odot\boldsymbol{W}=\boldsymbol{W}\text{diag}(\boldsymbol{\psi}_1)+\text{diag}(\boldsymbol{\psi}_2)\boldsymbol{W}-2(\boldsymbol{S}_{1}^{\intercal}\boldsymbol{S}_{2})\odot\boldsymbol{W},
\label{eq:meas_W}
\end{equation}
where the so-called connectivity matrix $\boldsymbol{W}$, which captures the $M$ available measurements, is defined as
\begin{equation}
  \color{black}
\boldsymbol{W}=\mleft[\begin{array}{@{\,}c|c@{\,}} \boldsymbol{1}_M\boldsymbol{1}_M^{\intercal}&\boldsymbol{1}_M\boldsymbol{1}_{{N}_2-M}^{\intercal}\\[0.5ex] \hline \boldsymbol{1}_{{N}_1-M}\boldsymbol{1}_M^{\intercal}&\boldsymbol{0}_{{N}_1-M}\boldsymbol{0}^{\intercal}_{{N}_2-M} \end{array} \mright],
\label{eq:connec_matrix}
\end{equation}
{\color{black} where $\boldsymbol{0}_{N}$ and $\boldsymbol{1}_N$ denote an all-zero/all-one column vector, respectively and the incomplete observations are captured by zeros in the bottom-right block of $\bm{W}$.}

\subsection{Problem Statement}
\label{sec:prob_stat}

With the aforementioned system  model in hand, we are ready to define the problem we seek to solve and, for the sake of context, discuss a particularly relevant \ac{SotA} method.
To that end, consider an augmented sensor location matrix carrying the positions of all landmark points in both bodies, such that we may write, in similarity to equation \eqref{eq:basic_model_one_body} 
\begin{equation}
\label{eq:big_S}
\!\boldsymbol{S}\!=\![\boldsymbol{S}_{1}|\boldsymbol{S}_{2}]\!=
\![\boldsymbol{Q}_{1}|\boldsymbol{Q}_{2}]\!
\mleft[
\begin{array}{@{\,}c@{\,}|@{\,}c@{\,}}
\boldsymbol{C}_1 & \boldsymbol{0}_{3\times {N_2}} \\
\hline
\boldsymbol{0}_{3\times {N_1}}&\boldsymbol{C}_2
\end{array}
\mright]\!\! + \![\boldsymbol{t}_{1}|\boldsymbol{t}_{2}]\!\mleft[
\begin{array}{@{\,}c|c@{\,}}
\boldsymbol{1}_{N_1}^{\intercal} & \boldsymbol{0}_{N_2}^{\intercal} \\[0.5ex]
\hline
\boldsymbol{0}_{N_1}^{\intercal}&\boldsymbol{1}_{N_2}^{\intercal}
\end{array}
\mright]\!,\!\!\!\!
\end{equation}
where $\boldsymbol{Q}_{i}$ and $\boldsymbol{t}_{i}$ denote the rotation matrix and translation vector of the $i$-th body respectively, while {\color{black}$\boldsymbol{0}_{3\times N}$ denotes an all-zero matrix.}

Under the egoistic assumption, $i.e.$ $\boldsymbol{S}_{1}=\boldsymbol{C}_{1}$ and $\boldsymbol{t}_{1} = \boldsymbol{0}_3$, however, equation \eqref{eq:big_S} reduces to
\begin{equation}
\label{eq:big_S_ego}
\boldsymbol{S}\!=\![\boldsymbol{S}_{1}\,|\,\boldsymbol{S}_{2}]\!=
\![\boldsymbol{I}\,|\,\boldsymbol{Q}]
\mleft[
\begin{array}{@{\,}c@{\,}|@{\,}c@{\,}}
\boldsymbol{C}_1 & \boldsymbol{0} \\
\hline
\boldsymbol{0}&\boldsymbol{C}_2
\end{array}
\mright]\!\! + \![\boldsymbol{0}\,|\,\boldsymbol{t}]\!\mleft[
\begin{array}{@{\,}c|c@{\,}}
\boldsymbol{1}_{N_1}^{\intercal} & \boldsymbol{0}_{N_2}^{\intercal} \\[0.5ex]
\hline
\boldsymbol{0}_{N_1}^{\intercal}&\boldsymbol{1}_{N_2}^{\intercal}
\end{array}
\mright]\!,\!\!\!\!
\end{equation}
where we have simplified the notation by omitting subscripts that can be inferred from context, which includes relabeling $\boldsymbol{Q} = \boldsymbol{Q}_{2}$ and $\boldsymbol{t} = \boldsymbol{t}_{2}$.

The problem addressed in this article -- namely, the attempt by rigid body 1 ($e.g.$, the truck in Fig. \ref{fig:sys}) to locate body 2 ($e.g.$, the car in Fig. \ref{fig:sys}) without support of infrastructure -- translates therefore to estimating, with basis on equations \eqref{eq:meas_full} and \eqref{eq:big_S_ego}, the rotation matrix $\boldsymbol{Q}$ and the translation vector $\boldsymbol{t}$, given perfect knowledge of the conformation matrix $\boldsymbol{C}_1$ (which implies exact knowledge of $\bm{D}_1$), possession of an incomplete estimate of the matrix $\bm{D}_{12}$ subject to noise, under the egoistic condition that $\boldsymbol{C}_2$ is unknown, and for a general case where $N_1 \neq N_2$.

\subsection{A Note on Related SotA}
\label{sec:Q_est}

To the best of our knowledge, the egoistic and generalized variation of the \ac{RBL} problem described above is original.
The closest related problem we are aware of is the one considered in \cite{PizzoICASSP2016}, which cast onto the description offered above would amount to the particular case where $N_1 = N_2$, combined with an idealistic assumption that $\boldsymbol{C}_2$ is perfectly and fully known.
In addition to these crucial distinctions, however, a critical error was made in \cite[Subsec. 3.2]{PizzoICASSP2016}, as demonstrated in Appendix A, which renders the approach thereby ineffective for the estimation of the translation vector $\boldsymbol{t}$.
In spite of the aforementioned error, the method in \cite{PizzoICASSP2016} partially inspired the contribution of our article to be introduced subsequently, such that it is useful to briefly revise in the sequel the portion of the method regarding the estimation of the rotation matrix $\boldsymbol{Q}$.

First, consider the $N\times N$ classic Sch{\"o}nberg \ac{DCM}, defined by \cite{Torgerson_1952} 
\vspace{-0.5ex}
\begin{equation}
\label{eq:J}
\boldsymbol{J} = \boldsymbol{I}-\frac{1}{N}\boldsymbol{1}\boldsymbol{1}^\intercal.
\vspace{-0.5ex}
\end{equation}

Taking into account incomplete observation and making use of the Sch{\"o}nberg \ac{DCM}, a projection matrix $\boldsymbol{P}$ can be formulated, written as
\begin{equation}
  \color{black}
\boldsymbol{P}=\mleft[\begin{array}{@{\,}c|c@{\,}} \boldsymbol{J}_M & \boldsymbol{0}_M\boldsymbol{0}_{N-M}^\intercal \\[0.5ex] \hline \boldsymbol{0}_{N-M}\boldsymbol{0}_M^\intercal & \boldsymbol{J}_{N-M} \end{array} \mright],
\end{equation}

Left- and right-multiplying a measured distance matrix by the projection matrix $\boldsymbol{P}$, and scaling the result by $-\frac{1}{2}$, as well as applying the connectivity matrix $\boldsymbol{W}$, yields
\begin{equation}
\begin{split}
\bar{\bm{D}}_{12}^{\odot 2} &= -\frac{1}{2}\boldsymbol{P}(\bm{D}_{12}^{\odot 2}\odot \boldsymbol{W})\boldsymbol{P}=(\boldsymbol{P}\boldsymbol{S}_1^\intercal\boldsymbol{S}_2\boldsymbol{P})\odot \boldsymbol{W}\\&=(\boldsymbol{C}_{1}^\intercal\boldsymbol{Q}\boldsymbol{C}_{2})\odot \boldsymbol{W},
\end{split}
\label{eq:D_tilde}
\end{equation}
where it was shown in \cite{PizzoICASSP2016} that due to the missing measurements only the visible measurements $\boldsymbol{C_{i,v}}$, $i.e.$, the first $M$ columns of the conformation matrix $\boldsymbol{C_i}$ are required.

To facilitate the formulation of a problem to estimate $\boldsymbol{Q}$, it will prove convenient to apply an \ac{OPP} onto equation \eqref{eq:D_tilde}, which under the assumption of perfect knowledge of $\boldsymbol{C}_{2}$ can be achieved by defining \cite{OPP_1966}
\begin{subequations}
\label{eq:right_mult}
\begin{equation}
\check{\bm{D}}_{12}^{\odot 2}\triangleq\bar{\bm{D}}_{12}^{\odot 2}\boldsymbol{C}_{2,v }^\dag=\boldsymbol{C}_{1,v }^\intercal\boldsymbol{Q},
\end{equation}
where
\begin{equation}
\boldsymbol{C}_{2,v}^\dag \triangleq \boldsymbol{C}_{2,v}^\intercal(\boldsymbol{C}_{2,v}\boldsymbol{C}_{2,v}^\intercal)^{-1}.
\end{equation}
\end{subequations}

Then, the relative rotation $\boldsymbol{Q}$ of body 2 with respect to the orientation of body 1 can be estimated by solving the problem
\begin{equation}
\hat{\boldsymbol{Q}}_{OPP}=\argmin_{\boldsymbol{Q}\in\mathbb{R}^{3\times 3}} ||\check{\bm{D}}_{12}^{\odot 2}-\boldsymbol{C}_{1,v}^\intercal\boldsymbol{Q}||_F^2,
\label{eq:OPP_opt}
\end{equation}
which can be obtained in closed form via \ac{SVD} of the matrix $\boldsymbol{C}_{1 }\check{\bm{D}}_{12}^{\odot 2}$.

In particular, the solution of problem \eqref{eq:OPP_opt} is given by \cite{PizzoICASSP2016}
\begin{subequations}
\label{eq:OPPsolution}
\begin{equation}
\hat{\boldsymbol{Q}}_{OPP}=\boldsymbol{U}\boldsymbol{V}^\intercal,
\end{equation}
with $\boldsymbol{U}$ and $\boldsymbol{V}$ such that
\begin{equation}
\boldsymbol{C}_{1,v }\check{\bm{D}}_{12}^{\odot 2} = \boldsymbol{U}\boldsymbol{\Sigma}\boldsymbol{V}^\intercal.
\end{equation}
\end{subequations}

While the solution to this problem is well known, it has to be noted that in order to use the orthogonal projection specific rank restrictions have to be fulfilled.
Specifically in our scenario we need to achieve $\text{rank}(\bar{C}_{2,v})=3$, where the rank can be generalized by the projection in \eqref{eq:D_tilde}, leading to  $\text{rank}(\bar{C}_{2,v})=M-1$.
This means that we need at least 4 links in order to perform the estimation and avoid singularities.

We emphasize that although it was assumed in \cite{PizzoICASSP2016} that both rigid bodies have the same number of landmark points ($e.g., N_1 = N_2$), the notion of a relative rotation \eqref{eq:big_S_ego} between two bodies is general, and therefore apply also to bodies of distinct shapes and/or different numbers of landmark points, as can be inferred from equation \eqref{eq:big_S_ego}.
In particular, by aligning the rotation matrix of the first rigid body with the {\color{black}Cartesian} coordinates, such that $\bm{Q}_1 = \bm{I}$, the orientation $\bm{Q}_2$ of the second body with respect to the first, becomes simply the relative rotation itself.
In other words, $\bm{Q}_1 = \bm{I} \Longrightarrow \bm{Q}_2 = \bm{Q}$, or more generally, $\bm{Q} = \bm{Q}_1^\intercal\cdot\bm{Q}_2$.

\vspace{-1ex}
\section{Egoistic Estimation of Translation Vectors between Arbitrarily-Shaped {\color{black}Rigid Bodies}}
\label{sec:prop}

The assumption of pre-existing knowledge of the conformation matrix $\bm{C}_2$, although typical of \ac{SotA} \ac{RBL} methods \cite{PizzoICASSP2016, Chen_2015}, is hard to meet in practical conditions.
In \ac{AD}-related \ac{V2X} applications, for instance, satisfying such assumption would require that a vehicle attempting to locate other vehicles in its vicinity is aware of their shapes, which is obviously impractical given the enormous diversity in vehicle models and their frequent updates.
In order to mitigate this problem, we propose in the following sections methods to estimate $\bm{t}$ and $\bm{Q}$, respectively, without the requirement that $\bm{C}_2$ is known.

In particular, in the following sections we first present the reconstruction of the \ac{EDM} in an egoistic scenario, followed by an \ac{MDS}-based egoistic translation estimation with corresponding performance evaluations.
Furthermore, a second estimator extending and correcting the \ac{SotA} method shown in \cite{PizzoICASSP2016} will be presented, which is shown to be more robust against incomplete observations. 

\subsection{MDS-based Egoistic Translation Estimation}
\label{sec:center_opt}

Let us start by pointing out that not knowing the conformation matrix $\bm{C}_2$ implicates not knowing the intra-distances matrix $\bm{D}_2$.
And while the reverse implication is not always true -- $i.e.$, in principle one could have knowledge of $\bm{D}_2$ but not $\bm{C}_2$ -- the assumption that $\bm{D}_2$ is also not available is consistent with egoistic principle followed in this article\footnote{Notice that an $N$-point \ac{3D} conformation matrix contains $3N$ entries, while the corresponding intra-distance matrix contains $N(N-1)/2$ distinct entries, such that the intra-distances data is larger than the conformation data for $N > 7$, which is a small number of points to define a rigid body in \ac{3D}.}.
In what follows, we therefore assume no knowledge of $\bm{D}_2$.

Under such conditions, the first problem at hand is one of matrix completion, and although several methods to solve such a problem exist \cite{Fang2012,Nguyen2019,Fan2024}, a number of which could be used, we here consider the simple and well-known Nystr\"om approximation method\footnotemark \cite{Williams_2001}.

\footnotetext{Although more sophisticated \ac{MC} exist, which could yield better performance, it will later be shown in Section \ref{sec:prop_journal} that our \ac{RBL} scheme offers additional robustness to incompletion, such that what is relevant to this article is the relative performances between the two proposed and the \ac{SotA} under a given \ac{MC} method.}

Applying the Nystr\"om approximation to the \ac{EDM} $\bm{D}$ from equation \eqref{eq:full_D} yields\footnote{The Nystr\"om approximation in general only works if the $\text{rank}(\bm{D}_1)\geq \text{rank}(\bm{D}_2)$, which means that the first body must have at least the same amount of sensors as the second body. If that condition is not satisfied, alternative matrix completion methods, $e.g.$ \cite{Fang2012,Nguyen2019,Fan2024}, may yield better results.} the following estimate of $\bm{D}_2$
\begin{equation}
\color{black}
\hat{\bm{D}}_2 \approx \mathbb{H}\big[\bm{D}_{12}^\intercal\bm{D}_1^{-1}\bm{D}_{12}\big],
\label{eq:NyAp}
\end{equation}
where $\mathbb{H}\big[\cdot\big]$ denotes a hollowing operator that enforces all elements of the diagonal matrix to be zero.

With the knowledge of the intra-distances matrix of the first body $\bm{D}_1$, calculated from the conformation matrix $\bm{C}_1$ itself, the measurements $\tilde{\bm{D}}_{12}$ corresponding to the distances between the two bodies, and the latter estimate $\hat{\bm{D}}_2$ of the intra-distances matrix corresponding to the second rigid body, the full sample \ac{EDM} corresponding to all distances within and between the two rigid bodies can be constructed as
\begin{equation}
\hat{\bm{D}}=\mleft[
\begin{array}{c|c}
\bm{D}_1 & \tilde{\bm{D}}_{12} \\
\hline\\[-2ex]
\tilde{\bm{D}}_{12}^{\intercal}&\mathbb{H}\big[\tilde{\bm{D}}_{12}^\intercal\bm{D}_1^{-1}\tilde{\bm{D}}_{12}\big]
\end{array}
\mright],
\label{eq:full_D_est}
\end{equation}
such {\color{black}that} an \ac{MDS}-based first estimate of the position of all sensors from both rigid bodies can be obtained as \cite{Torgerson_1952}
\begin{subequations}
\label{eq:MDSAlgo}
\begin{equation}
  \color{black}
\label{eq:S_MDS}
[\hat{\bm{S}}^{*}_{1}| \hat{\bm{S}}^{*}_{2}] =\bm{V} \bm{\Lambda}^{1/2},
\end{equation}
where $\bm{V}$ and $\bm{\Lambda}$ are the eigenvectors and eigenvalues of the corresponding double-centered \ac{EDM}, given by
\begin{equation}
\bar{\bm{D}} = \bm{V} \bm{\Lambda} \bm{V}^\intercal,
\end{equation}
\vspace{-1ex}
with
\begin{equation}
\bar{\bm{D}}= -\frac{1}{2}\boldsymbol{J}_{N_1+N_2}\hat{\bm{D}}^{\odot 2}\boldsymbol{J}_{N_1+N_2},
\end{equation}
where $\boldsymbol{J}_{N_1+N_2}$ is an $(N_1+N_2)$-point Sch{\"o}nberg \ac{DCM} build as described in equation \eqref{eq:J}.
\end{subequations}

The initial \ac{MDS} solution shown in equation \eqref{eq:S_MDS} can then be transformed to the reference frame of the first rigid body via a Procrustes transformation, namely
\begin{subequations}
\label{eq:Procrustes}
\begin{equation}
\label{eq:S2_est}
\hat{\bm{S}}_2 = \bm{R}^{*}\hat{\bm{S}}^{*}_{2} + \bm{t}^{*}\otimes \bm{1}_{N_2}^\intercal{\color{black},}
\end{equation}
\vspace{-1ex}
where
\begin{equation}
(\bm{R}^{*},\bm{t}^{*}) = \hspace{-3ex}\argmin_{\bm{R}\in\mathbb{R}^{3\times 3},\bm{t}\in\mathbb{R}^{3\times 1}} || \bm{C}_{1} - (\bm{R}\,\hat{\bm{S}}^{*}_{1} + \bm{t}\otimes \bm{1}_{N_1}^\intercal) ||_F,
\label{eq:MDS_Step}
\end{equation}
\end{subequations}
from which we then obtain the estimate
\begin{subequations}
\label{eq:FullS}
\begin{equation}
  \color{black}
\label{eq:S_est}
\hat{\bm{S}} = [\bm{C}_{1}| \hat{\bm{S}}_{2}],
\end{equation}
which substituted into equation \eqref{eq:big_S_ego} and using the relation $\boldsymbol{Q}_{2}\boldsymbol{C}_{2}=\hat{\boldsymbol{S}}_{2}\boldsymbol{J}_{N_2}$, yields
\vspace{-1ex}
\begin{equation}
  \color{black}
\label{eq:S_est_obj}
\begin{split}
\hat{\bm{S}}=\mleft[\boldsymbol{C}_1|(\bm{R}^{*}\hat{\bm{S}}^{*}_{2}\! +\! \bm{t}^{*}\!\otimes\!\bm{1}_{N_2}^\intercal)\boldsymbol{J}_{N_2}
\mright]\! +\! [\boldsymbol{0}|\boldsymbol{t}]\!\mleft[
\begin{array}{@{\,}c|c@{\,}}
\boldsymbol{1}_{N_1}^{\intercal} & \boldsymbol{0}_{N_2}^{\intercal} \\[0.5ex]
\hline
\boldsymbol{0}_{N_1}^{\intercal}&\boldsymbol{1}_{N_2}^{\intercal}
\end{array}
\mright].
\end{split}
\end{equation}
\end{subequations}

Utilizing the latter expression, the translation vector $\boldsymbol{t}$ can be found by solving the quadratic optimization program
\begin{subequations}
\label{tFinalMDS}
\begin{equation}
\hat{\boldsymbol{t}}=\argmin_{\boldsymbol{t}}||\boldsymbol{J}_{N_1+N_2}(\hat{\bm{S}}^\intercal\hat{\bm{S}}+\tfrac{1}{2}\hat{\bm{D}}^{\odot 2}\odot\hat{\boldsymbol{W}})\boldsymbol{J}_{N_1+N_2}||^2_F,
\label{eq:center_opt_IO}
\end{equation}
which can easily be solved by common optimization tools\footnote{Since the optimization problems presented in this paper are convex, their convergence to the global optimum is guaranteed, as known from convex optimization theory \cite{Boyd_Vandenberghe_2004}.
}, such as gradient descent or interior point methods \cite{Nocedal1999,Ruder2016}.

In the above, we have already integrated the notion of incomplete observations as described in Section \ref{sec:prob_stat}, by capturing incomplete observations via the erasure matrix
\begin{equation}
\hat{\boldsymbol{W}}=\mleft[\begin{array}{@{\,}c|c@{\,}}
\boldsymbol{I}_{N_1} & \boldsymbol{W} \\[0.5ex]
\hline
\boldsymbol{W}^{\intercal}&\boldsymbol{I}_{N_2}
\end{array}
\mright].
\end{equation}
\end{subequations}

The proposed method for the \ac{MDS}-based egoistic translation estimation without rigid body conformation knowledge, based on the range information between the two rigid bodies is therefore summarized in Algorithm \ref{alg:alg1}.

\begin{algorithm}[H]
\caption{: MDS Estimation of RBL Translation Vectors}
\label{alg:alg1}
\hspace*{\algorithmicindent}
\begin{algorithmic}[1]
\vspace{-0.9ex}
\Statex \hspace{-4ex} \textbf{Input:} Conformation matrix $\boldsymbol{C}_1$,  measurements $\tilde{\boldsymbol{D}}_{12}$. \vspace{-1.25ex}
\Statex \hspace{-4.4ex} \hrulefill
\Statex \hspace{-4ex}  \textbf{Output:} Translation vector estimate $\hat{\bm{t}}$; \vspace{-1.25ex}
\Statex \hspace{-4.4ex} \hrulefill
\State Construct $\boldsymbol{D}_1$ as the \ac{EDM} of $\boldsymbol{C}_1$, and $\hat{\boldsymbol{D}}$ via eq. \eqref{eq:full_D_est};
\State Obtain an estimate $\hat{\bm{S}}_2^*$ via MDS as per equation \eqref{eq:MDSAlgo};
\State Refine the latter estimate into $\hat{\bm{S}}_2$ via equation \eqref{eq:Procrustes};
\State Construct the stacked RBL estimate $\hat{\bm{S}}$ via equation \eqref{eq:FullS};
\State Solve eq. \eqref{tFinalMDS} to obtain the translation vector estimate $\hat{\bm{t}}$; 
\vspace{-2ex}
\end{algorithmic} 
\hspace*{\algorithmicindent}
\end{algorithm}
\vspace{-2ex}

\vspace{-2ex}
\subsection{Performance Evaluation}
\label{sec:res_MDS}
\vspace{-1ex}

\begin{figure}[H]
\centering
\includegraphics[width=\columnwidth]{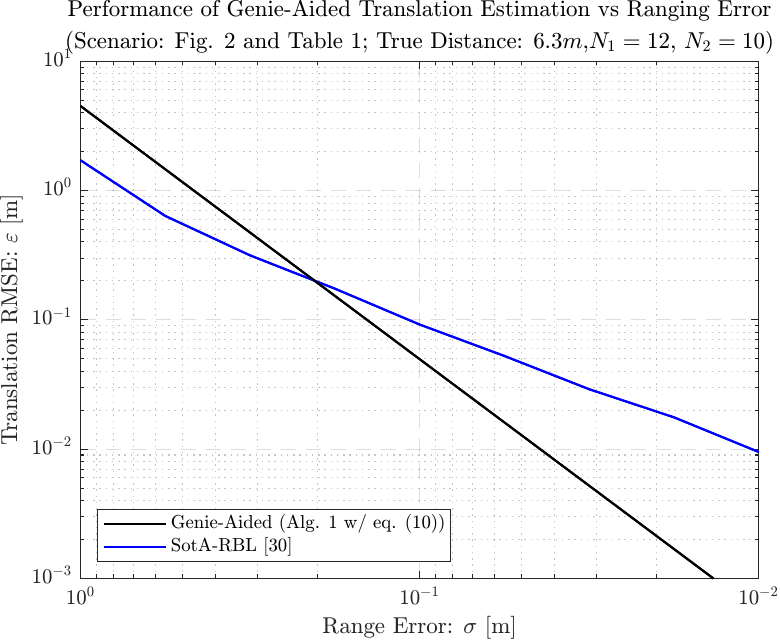}
\vspace{-5ex}
\caption{\ac{RMSE} of the translation estimate of the \acf{GA} proposed methods and the \ac{SotA}, over the range error $\sigma$.}
\label{fig:RMSE_Genie_MDS}
\end{figure}

In this section we offer simulation results illustrating the performance of the egoistic \ac{MDS}-based \ac{RBL} technique contributed above.
Since, to the best of our knowledge, no equivalent \ac{SotA} method exists for the egoistic set-up here considered, we first compare in Figure \ref{fig:RMSE_Genie_MDS} only the results of the estimation of the translation vector $\bm{t}$ via the non-egoistic method of \cite{Chen_2015}, against a variation of the proposed technique of Algorithm \ref{alg:alg1}, in which an estimate of $\bm{Q}$ obtained from \cite{Chen_2015} is used with the knowledge of the conformation matrix $\bm{C}_2$, such that equation \eqref{eq:S_est_obj} is effectively replaced by equation \eqref{eq:big_S_ego}.
For the sake of disambiguation, the proposed method will be dubbed Ego RBL, while the method modified by an externally fed conformation matrix is referred to as the ``Genie-Aided'' scheme.

\begin{table*}[ht]
\centering
\caption{Simulation Parameters}
\vspace{-2ex}
\begin{tabular}{|c|c|c|}
\hline
Reference frames & Translations & Rotations\\
\hline
\setlength{\arraycolsep}{2pt} 
$\boldsymbol{C}_1=\mleft[
\begin{array}{*{12}{c}}
-1.25 & 1.25 & -1.25 & 1.25 & -1.25 & 1.25 & -1.25 & 1.25 & -1.25 & 1.25 & -1.25 & 1.25 \\
-4 & -4 & -4 & -4 & 0 & 0 & 0 & 0 & 4 & 4 & 4 & 4 \\
0.5 & 0.5 & 1 & 1 & 1 & 1 & 4 & 4 & 4 & 4 & 0.5 & 0.5 \\
\end{array}
\mright]$ & $\boldsymbol{t}_1=[0, 0, 0]^\intercal$ & $[\psi_1,\theta_1,\phi_1]=[0^\circ,0^\circ,0^\circ]$\\
\setlength{\arraycolsep}{2.5pt} 
$\boldsymbol{C}_2=\mleft[
\begin{array}{*{12}{c}}
-1 & 1 & -1 & 1 & -1 & 1 & -1 & 1 & -1 & 1 \\
2 & 2 & 1 & 1 & -1 & -1 & -2 & -2 & 0 &0  \\
1 & 1 & 1.5 & 1.5 & 1.5 & 1.5 & 1 & 1 & 0.5 & 0.5 \\
\end{array}
\mright]$
&
\begin{tabular}{@{}c@{}c@{}}
$\boldsymbol{t}_2=\boldsymbol{t}=[7,3, 0.5]^\intercal$\Tstrut
\vspace{1ex}
\end{tabular}
&
\begin{tabular}{@{}c@{}c@{}}
$[\psi_2,\theta_2,\phi_2]=[10^\circ,20^\circ,45^\circ]$
\end{tabular}\\
\hline
\end{tabular}
\label{table::parameters}
\vspace{-3ex}
\end{table*}

To analyze the behavior and performance of the two proposed methods, we choose the \ac{RMSE} as our metric for comparison.
Thus, we need to define the \ac{RMSE} for the translation vector as

\vspace{-5ex}
\begin{figure}[H]
\centering
\vspace{1ex}
\includegraphics[width=\columnwidth]{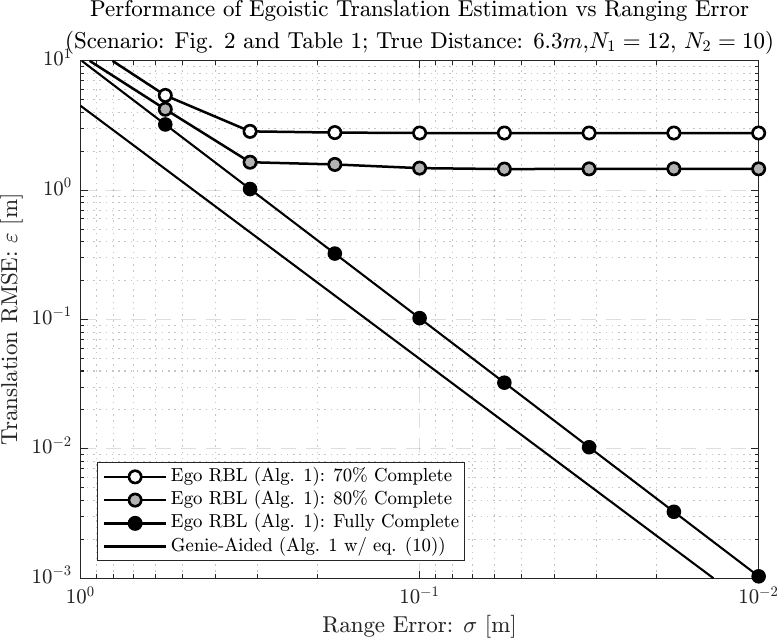}
\vspace{-5ex}
\caption{\ac{RMSE} of the translation estimate of the proposed \ac{MDS}-based method for different levels of available information, over the range error $\sigma$.}
\label{fig:RMSE_IO_Conf}
\end{figure}
\vspace{-2ex}
\vspace{-1ex}
\begin{equation}
\varepsilon = \frac{1}{\sqrt{K}}\bigg(\sum_{k=1}^{K}|\hat{\boldsymbol{t}}^{(k)}-{\boldsymbol{t}}|_2^2\bigg)^{\!\!\frac{1}{2}},
\vspace{-1ex}
\end{equation}
where $\hat{\boldsymbol{t}}$ is the estimated translation vector and $K=10^3$ is the number of Monte-Carlo simulations\footnotemark.

\footnotetext{The algorithms are implemented in MATLAB, with the minimization problems solved using the CVX optimization package \cite{cvx,gb08}.}

The considered scenario is represented by two rigid bodies as shown in Fig. \ref{fig:sys}, where the parameters, including translation, rotation, as well as the original reference frames of the rigid bodies, are summarized in Table \ref{table::parameters} and will be consistently utilized hereafter, unless when otherwise stated.
Our first results are given in Fig. \ref{fig:RMSE_Genie_MDS} comparing the translation estimates in a non-egoistic scenario, between the proposed method and the \ac{SotA} in \cite{Chen_2015} respectively in terms of \ac{RMSE} in meter over the range error\footnote{Note that the range error is not equivalent to the exact error in meters but rather the error used in the noise calculations given in \eqref{eq:sqr_meas}.} $\sigma$.

It can be observed in Figure \ref{fig:RMSE_Genie_MDS} that the proposed \ac{MDS}-based method outperforms the \ac{SotA} of \cite{Chen_2015} for range errors below $20$ cm, which is well within the typical accuracy of sensing technology used in Automotive Industry \cite{MalekianSenJ2018}.
In turn, in Figure \ref{fig:RMSE_IO_Conf}, it is found that the proposed egoistic scheme for a fully complete scenario achieves a performance very close to that of the Genie-Aided variation, confirming that the contributed (egoistic) \ac{MDS}-based method is capable of handling the practical condition that no prior knowledge on the shape of the target object is available{, which makes the method suitable for real-world applications. }

Finally, further simulations are performed to evaluate the performance of Algorithm \ref{alg:alg1} in systems under different levels of incompletion.
The results, also shown in Figure \ref{fig:RMSE_IO_Conf}, illustrate the impact of incomplete information on the accuracy \ac{RBL}.
In particular, it is seen that performance degrades quickly, leading to rather high error floors as the available information goes from fully complete to $80\%$, to $70\%$ complete\footnote{The percentage of completion is computed based on the number of zeros and non-zeros in the connectivity matrix $\bm{W}$ given in equation \eqref{eq:connec_matrix}{\color{black}, by varying $M$}.}.

\vspace{-3ex}
\subsection{Matrix Completion aided Localization}
\label{sec:prop_MCRB}
\vspace{-.5ex}

As seen in Figure \ref{fig:RMSE_IO_Conf}, incompleteness in the distance measurement matrix significantly harms the estimation performance.
To counter this harming effect, \acf{MC} methods can be used.
We emphasize, however, that performing \ac{MC} directly over the entire \ac{EDM} $\boldsymbol{D}$, which is a hollow symmetric matrix, usually yields poor results.
Instead, we apply \ac{MC} to fill the missing elements of the partially observed measurement matrix $\tilde{\bm{D}}_{12}$, and then construct the estimate of $\hat{\bm{D}}$ via Nystr\"om approximation, via equation \eqref{eq:full_D_est}.

Well known and high-performing matrix completion algorithms include the simple rank enforcing algorithm \cite{RankEDM}, the OptSpace algorithm \cite{OptSpace}, the \ac{SI} method \cite{Mazumder_2010}, and the accelerated and inexact Soft-Inpute (AIS-Impute) approach of \cite{Yao_2019}.  
Among these techniques, we select OptSpace \cite{OptSpace} due to its trade-off between complexity and performance, but obviously any other suitable method can be used.

\begin{figure}[H]
\centering
{{\includegraphics[width=\columnwidth]{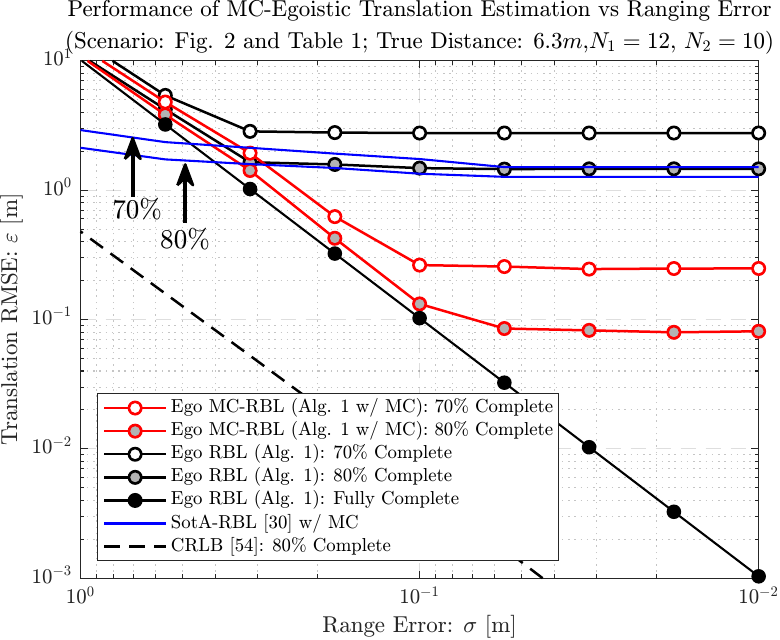}}}
\vspace{-5ex}
\caption{\ac{RMSE} of the translation estimate of the proposed method aided by matrix completion for different levels of available information, over the range error $\sigma$.}
\label{fig:RMSE_MC_Conf}
\vspace{-1ex}
\end{figure}

Figure \ref{fig:RMSE_MC_Conf} illustrates the impact of matrix completion onto the egoistic MDS-based translation estimation technique proposed in Subsection {\ref{sec:center_opt}}, under the same conditions of Figure \ref{fig:RMSE_IO_Conf}.
It can be observed that indeed matrix completion can lower the error floors observed previously, approximately by a factor of 10.
Additionally, the performance of the method of \cite{Chen_2015} enhanced via matrix completion is shown (for $70\%$ and $80\%$ available information), demonstrating that the \ac{SotA} approach is not improved by matrix completion.
{While the performance of the proposed method is still inferior to that of the fully complete case, it is seen that \ac{MC} can significantly alleviate the harmful effect of incompleteness in the observations compared to the \ac{SotA}.
Nevertheless, there is still room for improvement, as shown by the \ac{CRLB}, which was calculated following the formulation described in \cite{fuehrling2025fundamentallimits}.}

\vspace{-1ex}
\section{Robust Egoistic Translation Estimation}
\label{sec:prop_journal}

In spite of its attractive features, in particular the self-reliant framework in terms of not requiring prior knowledge on the conformation of the target object, and the ability to handle rigid bodies of arbitrary shapes with distinct numbers of landmark points, the technique proposed and evaluated in Section \ref{sec:prop} has one limitation that can be improved upon.
Although robustness to incompleteness in $\hat{\bm{D}}$ can be partially alleviated by the incorporation of \ac{MC}, as discussed above, the method itself offers no particular (additional) means to increase robustness.

\vspace{-2ex}

\subsection{MDS-based Egoistic Translation Estimation}
\vspace{-1ex}
 In light of the above, we introduce a second method, which extends the corrected version of the translation distance estimator of \cite{PizzoICASSP2016}, detailed in Appendix \ref{Corrected_Est}, which is designed to enable the robust estimation of the translation vector $\bm{t}$ in an egoistic setup.
To that end, first consider the corrected translation distance estimator given in equation \eqref{eq:rel_tra_est} of Appendix \ref{Corrected_Est}, which for the sake of convenience is reproduced below
\vspace{-1.5ex}
\begin{eqnarray}
\label{eq:rel_tra_estAgain}
\hat{t}=\frac{1}{N^2}||\bm{D}_{12}||_F^2-\frac{1}{N}\sum^N_{n=1}(||\boldsymbol{c}_{1,n}||^2_2+||\boldsymbol{c}_{2,n}||^2_2)+&&\\
&&\hspace{-49ex}\vspace{-1ex}
\text{\raisebox{-20pt}{\rotatebox{90}{$\text{missing in}\atop \text{\cite[Sec. 3.2]{PizzoICASSP2016}}$}}}
\begin{cases}
\displaystyle
+\frac{2}{N^2}\bm{1}^\intercal(\boldsymbol{C}_{1}^\intercal\boldsymbol{Q}_{1}^\intercal\boldsymbol{Q}_{2}\boldsymbol{C}_{2}+\boldsymbol{C}_{1}^\intercal\boldsymbol{Q}_{1}^\intercal\boldsymbol{t}_{2}\bm{1}^{\intercal}+\bm{1}\boldsymbol{t}_{1}^\intercal \boldsymbol{Q}_{2}\boldsymbol{C}_{2})\bm{1}\\
\displaystyle
-\frac{1}{N}\sum_{i=1}^N(2\boldsymbol{c}_{i,1}^\intercal\boldsymbol{Q}_1^\intercal\boldsymbol{t}_1+2\boldsymbol{c}_{i,2}^\intercal\boldsymbol{Q}_2^\intercal\boldsymbol{t}_2).\nonumber
\end{cases}
\end{eqnarray}

{\color{black}However}, since in the egoistic scenario the conformation matrix of the second rigid body is unknown, equation \eqref{eq:Corrected_Est_S} has to be considered, given by
\vspace{-1ex}
\begin{align}
\label{eq:sota_corrected_S}
\hat{t}&=\frac{-1}{N}\!\sum^N_{n=1}(||\boldsymbol{s}_{1,n}||^2_2\!+\!||\boldsymbol{s}_{2,n}||^2_2)\!+\!||\boldsymbol{t}_{1}||^2_2\!+\!||\boldsymbol{t}_{2}||^2_2\!+\!\frac{1}{N^2}||\bm{D}_{12}||_F^2\nonumber\\
&+\!\frac{2}{N^2}\bm{1}^\intercal(\boldsymbol{C}_{1}^\intercal\boldsymbol{Q}_{1}^\intercal\boldsymbol{Q}_{2}\boldsymbol{C}_{2}\!+\!\boldsymbol{C}_{1}^\intercal\boldsymbol{Q}_{1}^\intercal\boldsymbol{t}_{2}\bm{1}^{\intercal}\!+\!\bm{1}\boldsymbol{t}_{1}^\intercal \boldsymbol{Q}_{2}\boldsymbol{C}_{2})\bm{1}.\!\!
\end{align}

By applying the same knowledge, relations as before, {\color{black}$i.e.$, $N_1\neq N_2$, the specific parameters shown in Table \ref{table::parameters}, and the egoistic relation $\boldsymbol{Q}_{2}\boldsymbol{C}_{2}=\hat{\boldsymbol{S}}_{2}\boldsymbol{J}_{N_2}$,} \eqref{eq:sota_corrected_S} can be rewritten to
\begin{eqnarray}
\label{eq:Ego_est_t}
\hat{t}=\boldsymbol{t}^\intercal\boldsymbol{t}=\frac{-1}{N_1}\!\sum^{N_1}_{n=1}(||\boldsymbol{s}_{1,n}||^2_2)\!+\!\frac{-1}{N_2}\!\sum^{N_2}_{n=1}(||\boldsymbol{\hat{s}}_{2,n}||^2_2)\!+\!||\boldsymbol{t}||^2_2&&\\
&&\hspace{-50ex}+\!\frac{2}{N_1N_2}(||\bm{D}_{12}||_F^2+\bm{1}_{N_1}^\intercal(\boldsymbol{C}_{1}^\intercal{\color{black}\hat{\boldsymbol{S}}_2} J_{N_2}\!+\!\boldsymbol{C}_{1}^\intercal\boldsymbol{t}_{2}\bm{1}_{N_2}^{\intercal}\!)\bm{1}_{N_2}).\!\!\nonumber
\end{eqnarray}

Since the objective of our proposed method is to estimate the translation vector instead of the distance only, the problem requires the isolation of $\boldsymbol{t}$ in \eqref{eq:Ego_est_t} to be able to construct a corresponding optimization problem.
The translation vector $\boldsymbol{t}$ is contained in the cross terms, which can be isolated by using vectorization \cite{Macedo_2013}, with the corresponding term written as
\begin{equation}
  \label{eq:isolate_t}
\!\!\!
\frac{2}{N_1N_2}\boldsymbol{1}_{N_1}^\intercal(\boldsymbol{C}_{1}^\intercal\boldsymbol{t}_{2}\boldsymbol{1}_{N_2}^{\intercal})\boldsymbol{1}_{N_2}=\frac{2}{N_1N_2}\boldsymbol{1}_{N_1N_2}^\intercal(\boldsymbol{1}_{N_2}\otimes\boldsymbol{C}_{1}^\intercal)\boldsymbol{t}.
\end{equation}

{\color{black}By rearranging equation \eqref{eq:Ego_est_t} and the isolation of $\boldsymbol{t}$ by plugging in equation \eqref{eq:isolate_t}}, the estimator can be reformulated and simplified to
\begin{equation*}
\nonumber
\boldsymbol{t}^\intercal\boldsymbol{t}=\tfrac{-1}{N_1}\!\sum^{N_1}_{n=1}(||\boldsymbol{s}_{1,n}||^2_2)\!-\!\tfrac{1}{N_2}\!\sum^{N_2}_{n=1}(||\boldsymbol{\hat{s}}_{2,n}||^2_2)\!+\!\boldsymbol{t}^\intercal\boldsymbol{t}\!+\!\tfrac{1}{N_1N_2}||\bm{D}_{12}||_F^2
\vspace{-1ex}
\end{equation*}
\begin{equation}
\!\!+\tfrac{2}{N_1N_2}\boldsymbol{1}_{N_1N_2}^\intercal(\boldsymbol{1}_{N_2}\otimes\boldsymbol{C}_{1}^\intercal)\boldsymbol{t}+\!\tfrac{2}{N_1N_2}\bm{1}_{N_1}^\intercal(\boldsymbol{C}_{1}^\intercal\hat{S}_2 J_{N_2}\!)\bm{1}_{N_2}.
\end{equation}


Next, rearranging the equation with its only unknown being $\boldsymbol{t}$, the resulting problem can be written as
\begin{subequations}
\label{eq:final_est_eq}
\begin{equation}
    0=\boldsymbol{a}\boldsymbol{t}+b,
\end{equation}
\vspace{-1ex}
where
\begin{equation}
    \label{eq:final_est_eq_a}
    \bm{a}\triangleq\frac{2}{N_1N_2}\boldsymbol{1}_{N_1N_2}^\intercal(\boldsymbol{1}_{N_2}\otimes\boldsymbol{C}_{1}^\intercal),
\end{equation}
\vspace{-4ex}
\begin{align}
    \label{eq:final_est_eq_b}
    {\color{black}b}\triangleq&\frac{-1}{N_1}\!\sum^{N_1}_{n=1}(||\boldsymbol{s}_{1,n}||^2_2)\!+\!\frac{-1}{N_2}\!\sum^{N_2}_{n=1}(||\boldsymbol{\hat{s}}_{2,n}||^2_2)\!+\!\frac{1}{N_1N_2}||\bm{D}_{12}||_F^2 \nonumber\\
&+\!\frac{2}{N_1N_2}\bm{1}_{N_1}^\intercal(\boldsymbol{C}_{1}^\intercal\hat{S}_2 J_{N_2}\!)\bm{1}_{N_2}.
\end{align}
\end{subequations}

Since the unknown variable is not a scalar but a vector, the problem is underdetermined and does not have a closed form solution.
Thus, before formulating an optimization problem, a constraint has to be designed that resolves this issue, additionally adding robustness to the problem.
Therefore, a constraint can be added to the problem that minimizes the distance of an artificial distance matrix found by the optimization itself, compared to the measured one.
To that extend, an artificial distance matrix can be formed via equation \eqref{eq:meas_full}, written as
\begin{subequations}
\begin{equation}
\label{eq:aux_constr}
\boldsymbol{D}_{12}^*=\boldsymbol{\psi}_1\boldsymbol{1}_N^{\intercal}+\boldsymbol{1}_N\boldsymbol{\psi}_2^{\intercal}-2\boldsymbol{S}_{1}^{\intercal}\boldsymbol{S}_{2}^*,
\end{equation}
with $\boldsymbol{\psi}_i=\big[||\boldsymbol{s}_{1,i}||_2^2,\cdots,||\boldsymbol{s}_{N,i}||_2^2\big]^{\intercal}$, and $\boldsymbol{t}$ contained in
\begin{equation}
\boldsymbol{S}_{2}^*=\boldsymbol{Q}\boldsymbol{C}_{2}+\boldsymbol{t}\boldsymbol{1}_N^{\intercal}=\hat{\boldsymbol{S}}_{2}\boldsymbol{J}_{N_2}+\boldsymbol{t}^*\boldsymbol{1}_N^{\intercal},
\end{equation}
\end{subequations}
which uses the estimated centered $\hat{\boldsymbol{S}}_2$ from the \ac{MDS} solution and the variable to optimize, $i.e.$, the translation vector $\boldsymbol{t}$.

Adding the constraint to the objective, the constraint optimization problem can be written as
\begin{subequations}
\vspace{-1ex}
\label{eq:opt_corr}
\begin{equation}
    \label{eq:opt_corr_obj}
\min_{\boldsymbol{t}} |\boldsymbol{a}\boldsymbol{t}+b|,
\end{equation}
\vspace{-1ex}
s.t.
\begin{equation}
    \label{eq:opt_corr_constr}
    || (\boldsymbol{D}_{12}^*- \hat{\boldsymbol{D}}_{12})\odot\bm{W}||^2_F\leq \epsilon,
\end{equation}
\noindent where $\boldsymbol{a}$ and $b$ denote the parts of equation defined in \eqref{eq:final_est_eq_a} and \eqref{eq:final_est_eq_b} respectively, {and $\boldsymbol{D}_{12}^*$ denotes the auxiliary variable representing the reconstructed distance matrix from equation \eqref{eq:meas_full}, as described in equation \eqref{eq:aux_constr}.
Furthermore, $\epsilon=0.01$ represents a small noise regularization term introduced to ensure numerical stability and prevent singularities during optimization.
The value of $\epsilon$ was chosen empirically after testing a range of values from $0.001$ to $0.1$ and observing negligible differences in performance. 
}

\end{subequations}

As before, the problem can be solved via conventional optimization tools, such as gradient descent or interior point methods \cite{Nocedal1999,Ruder2016}, and its robustness compared to the method described in Section \ref{sec:prop} can be justified by comparing the two problems itself.
In particular, while in the first method the impact of missing measurements in equation \eqref{tFinalMDS} can only be compensated by the terms depending on $\bm{t}$, in the second problem, zeros in the distance matrix do not have such a big impact on the estimate.
This is observable in the constraint shown in equation \eqref{eq:opt_corr_constr}, where the zeros are common in both required matrices, and in the objective shown in equation \eqref{eq:opt_corr_obj} itself, where the Frobenius norm of the distance matrix is scaled, which does not harm the estimate.

The second proposed method, which for the sake of disambiguation is therefore dubbed Robust Ego RBL, is summarized in Algorithm \ref{alg:alg2}.

\begin{algorithm}[H]
\caption{: Robust Egoistic Translation Estimation}
\label{alg:alg2}
\hspace*{\algorithmicindent}
\begin{algorithmic}[1]
\vspace{-0.9ex}
\Statex \hspace{-4ex} \textbf{Input:} Conformation matrix $\boldsymbol{C}_1$,  measurements $\tilde{\boldsymbol{D}}_{12}$, hyperparameter $\epsilon$. \vspace{-1.25ex}
\Statex \hspace{-4.4ex} \hrulefill
\Statex \hspace{-4ex}  \textbf{Output:} Translation vector estimate $\hat{\bm{t}}$; \vspace{-1.25ex}
\Statex \hspace{-4.4ex} \hrulefill
\State Construct $\boldsymbol{D}_1$ as the \ac{EDM} of $\boldsymbol{C}_1$ and $\hat{\boldsymbol{D}}$ via eq. \eqref{eq:full_D_est};
\State Obtain an estimate $\hat{\bm{S}}_2^*$ via MDS as per equation \eqref{eq:MDSAlgo};
\State Refine the latter estimate into $\hat{\bm{S}}_2$ via equation \eqref{eq:Procrustes};
\State Construct the stacked RBL estimate $\hat{\bm{S}}$ via equation \eqref{eq:FullS};
\State Solve eq. \eqref{eq:opt_corr} to obtain the translation vector estimate $\hat{\bm{t}}$; 
\vspace{-2ex}
\end{algorithmic} 
\hspace*{\algorithmicindent}
\end{algorithm}

\vspace{-2ex}

\subsection{Performance Evaluation}

\begin{figure}[H]
\centering
{{\includegraphics[width=\columnwidth]{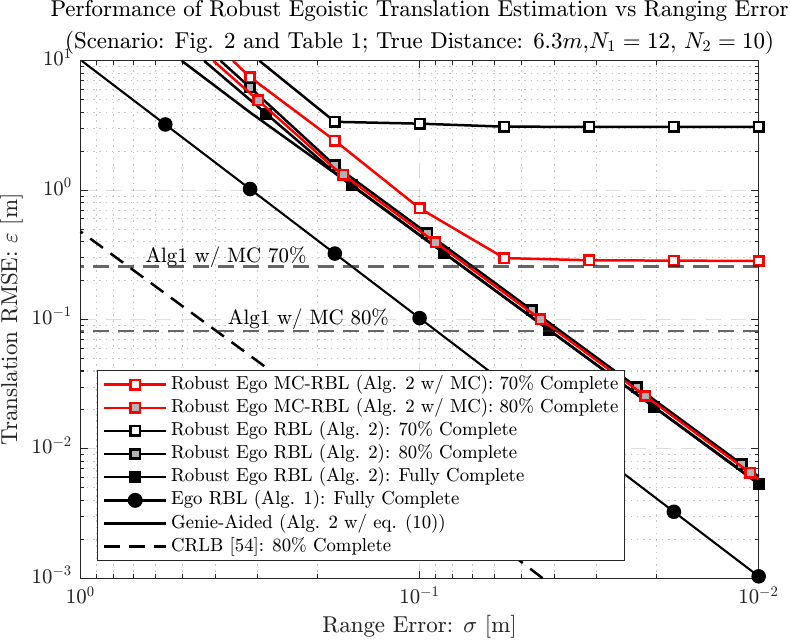}}}
\vspace{-4ex}
\caption{\ac{RMSE} of the translation estimate of the proposed robust method compared to the \ac{MDS}-based method aided by matrix completion for different levels of available information, over the range error $\sigma$.}
\label{fig:RMSE_MC_J}
\end{figure}

To evaluate the performance of the robust method, the scenario presented in {Section} \ref{sec:res_MDS} is revisited, comparing the two proposed methods with a focus on the robustness to incomplete observations, as described in {Section \ref{sec:prop_MCRB}}.
The corresponding results are presented in Figure \ref{fig:RMSE_MC_J}.
The first set of results compare the translation estimates between both proposed methods in a non-egoistic \ac{GA} scenario, and the egoistic scenario respectively in terms of \ac{RMSE}.
It can be observed that while the robust method performs worse than the \ac{MDS}-based method, since the latter more effectively preserves the global geometry and relative distances in the data, the performance of the egoistic robust method is nearly identical to the \ac{GA} variation.

Also illustrated in Figure \ref{fig:RMSE_MC_J}, simulations are performed again for different levels of available information, also incorporating the matrix completion aided Algorithm \ref{alg:alg2}.
While for $80\%$ available information the performance is close to the fully complete scenario, for $70\%$ available information the method results in an error floor that can be improved by using matrix completion.

Additionally, as shown by the dashed lines that correspond to the error floors of Algorithm \ref{alg:alg1} at different levels of completeness, for $70\%$ the two methods perform similar, while for $80\%$ Algorithm \ref{alg:alg2} outperforms the \ac{MDS}-based method in low range error regimes.
{Overall, the results confirm that the proposed robust egoistic translation estimation method can effectively handle incomplete observations, especially when aided by matrix completion, making it suitable for practical applications where data may be missing or incomplete.}

\vspace{-2ex}
\section{Egoistic Rotation Matrix Estimation}
\label{sec:Q_C_est}

Finally, we offer an egoistic rotation matrix estimator which works solely with distance measurements without requiring knowledge of the conformation matrix $\bm{C}_2$, and which is entirely complementary to the latter contributions in so far as it also does not require translation vector estimates.

In particular, with the initial location estimate matrix $\hat{\bm{S}}_2$ obtained via equation \eqref{eq:S2_est} in hands, we seek to obtain an estimate $\hat{\bm{Q}}$ of the rotation matrix $\bm{Q}$ of the target rigid body via a procedure similar to that described in Subsection \ref{sec:Q_est}.
Before we proceed, let us emphasize that, in principle, $\hat{\bm{Q}}$ can be extracted from the relation
\begin{equation}
\label{eq:Q_est_eig}
(\hat{\bm{S}}_2\bm{J}_{N_2})(\hat{\bm{S}}_2\bm{J}_{N_2})^\intercal = \bm{Q} \bm{\Lambda} \bm{Q}^\intercal
= \bm{Q} \boldsymbol{C}_{2}\boldsymbol{C}_{2}^\intercal \bm{Q}^\intercal,
\end{equation}
where we used $\boldsymbol{Q}_{2}\boldsymbol{C}_{2}=\hat{\boldsymbol{S}}_{2}\boldsymbol{J}_{N_2}$ in the last equality.

Notice, however, that the eigenvalue decomposition in equation \eqref{eq:Q_est_eig} is such that the eigenvectors are ordered according to the magnitude of the corresponding eigenvalues, which in turn relate to the largest orthogonal dimensions of the body itself \cite{jolliffe2002principal,hastie2009elements}.
Consequently, the columns of the estimate obtained from equation \eqref{eq:Q_est_eig} may be swapped for rigid bodies without distinctly different length, width and height, which may lead to large estimation errors.
In order to circumvent this issue, we instead propose the method described below.

First, let us return to equation \eqref{eq:D_tilde}, but this time accounting for the fact that $\bm{S}_1$ and $\bm{S}_2$ can have different numbers $N_1$ and $N_2$ of landmark points, such that
\vspace{-0.5ex}
\begin{equation}
    \label{eq:d_bar}
\bar{\bm{D}}_{12}^{\odot 2}= -\frac{1}{2}\boldsymbol{J}_{N_1}\bm{D}_{12}^{\odot 2}\boldsymbol{J}_{N_2}=\boldsymbol{C}_{1}^\intercal\boldsymbol{Q}\boldsymbol{C}_{2},
\vspace{-0.5ex}
\end{equation}
which{\color{black},} if left-multiplied by the pseudo-inverse of $\boldsymbol{C}_{1}^\intercal$ yields
\vspace{-0.5ex}
\begin{subequations}
    \label{eq:D_check}
\begin{equation}
\label{eq:left_mult}
\check{\bm{D}}_{12}^{\odot 2}\triangleq\boldsymbol{C}_{1}^\dag\bar{\bm{D}}_{12}^{\odot 2}= \bm{Q}\boldsymbol{C}_{2},
\vspace{-0.5ex}
\end{equation}
where
\vspace{-0.5ex}
\begin{equation}
\boldsymbol{C}_{1}^\dag \triangleq (\boldsymbol{C}_{1}\boldsymbol{C}_{1}^\intercal)^{-1}\boldsymbol{C}_{1}.
\vspace{-0.5ex}
\end{equation}
\end{subequations}

Then, squaring equation \eqref{eq:left_mult} yields
\vspace{-0.5ex}
\begin{equation}
\check{\bm{D}}_{12}^{\odot 2}\check{\bm{D}}_{12}^{\odot 2 \intercal}= \bm{Q}\boldsymbol{C}_{2}\boldsymbol{C}_{2}^\intercal\bm{Q}^\intercal = \bm{Q}\boldsymbol{\Lambda}\bm{Q}^\intercal,
\vspace{-0.5ex}
\end{equation}
leading to the optimization problem
\vspace{-0.5ex}
\begin{equation}
\hat{\boldsymbol{Q}}=\argmin_{\boldsymbol{Q}} ||\check{\bm{D}}_{12}^{\odot 2}\check{\bm{D}}_{12}^{\odot 2 \intercal}-\bm{Q}\bm{\Lambda}\bm{Q}^\intercal||_F^2.
\label{eq:OPP_opt_2}
\vspace{-0.5ex}
\end{equation}

We emphasize that although the solution of problem \eqref{eq:OPP_opt_2} can be easily obtained via common optimization theory tools \cite{Nocedal1999,Ruder2016}, the result can also be severely degraded by the order of the eigenvalues in $\bm{\Lambda}$.
Fortunately, however, in \ac{3D} there are only 6 distinct permutations of $\bm{\Lambda}$, such that the solution with the permutation that yields the smallest objective can be estimated as the correct one.

The proposed egoistic rotation matrix estimation method is summarized in Algorithm \ref{alg:alg3} for  rotation estimation without rigid body conformation knowledge, based on the range information between the two rigid bodies, independent of the translation vector estimate.

\vspace{-1ex}
\begin{algorithm}[H]
    \caption{: Rotation Estimation}
    \label{alg:alg3}
    \hspace*{\algorithmicindent}
    \begin{algorithmic}[1]
    \vspace{-0.9ex}
    \Statex \hspace{-4ex} \textbf{Input:} Conformation matrix $\boldsymbol{C}_1$,  measurements $\tilde{\boldsymbol{D}}_{12}$. \vspace{-1.25ex}
    \Statex \hspace{-4.4ex} \hrulefill
    \Statex \hspace{-4ex}  \textbf{Output:} Rotation matrix estimate $\hat{\bm{Q}}$; \vspace{-1.25ex}
    \Statex \hspace{-4.4ex} \hrulefill
    \State Construct $\boldsymbol{D}_1$ as the \ac{EDM} of $\boldsymbol{C}_1$ and $\hat{\boldsymbol{D}}$ via eq. \eqref{eq:full_D_est};
    \State Obtain an estimate $\hat{\bm{S}}_2^*$ via MDS as per equation \eqref{eq:MDSAlgo};
    \State Refine the latter estimate into $\hat{\bm{S}}_2$ via equation \eqref{eq:Procrustes};
    \State Construct the double-centered distance matrix $\bar{\bm{D}}_{12}^{\odot 2}$ via equation \eqref{eq:d_bar};
    \State Refine the latter into $\check{\bm{D}}_{12}^{\odot 2}$ via equation \eqref{eq:D_check};
    \State Solve eq. \eqref{eq:OPP_opt_2} to obtain the rotation matrix estimate $\hat{\bm{Q}}$; 
    \vspace{-2ex}
    \end{algorithmic} 
    \hspace*{\algorithmicindent}
\end{algorithm}

\vspace{-4ex}
\subsection{Performance Evaluation}

Next, {\color{black} we} seek to evaluate the performance of both the translation vector and rotation matrix in terms of their impact on the overall pose estimation of the target rigid body.
In doing so, we also rule out simply computing the \ac{RMSE} of all the landmark points of the target rigid both, so as to keep true to the egoistic characteristic of the contribution, which is not to rely on knowledge of the shape of the target.

In order to circumvent this challenge and capture the errors due to {\color{black}the independent translation and rotation estimation} jointly, without the explicit knowledge of the conformation matrix, we propose to model each rigid body as unit vector with its origin at the location of body, and its direction as determined by its rotation matrix.

\begin{figure}[H]
\centering
\includegraphics[width=\columnwidth]{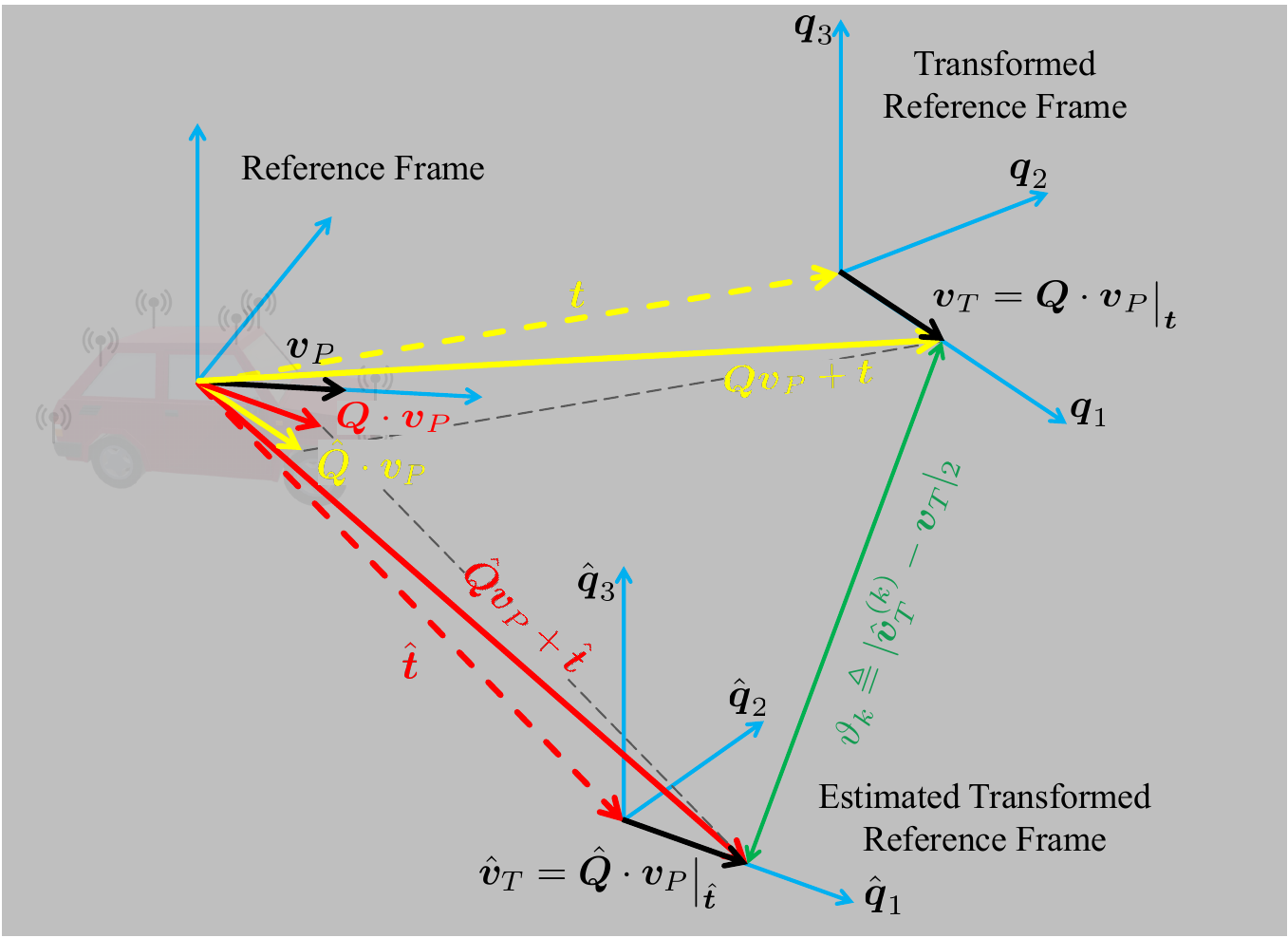}
\vspace{-4ex}
\caption{Illustration of a rigid body modeled as a unit vector $\bm{v}_P$, with the true transformed vector $\bm{v}_T$ and its estimate $\hat{\bm{v}}_T$.}
\label{fig:Sem_RBL_Model}
\vspace{-4ex}
\end{figure}

To clarify, referring to Figure \ref{fig:Sem_RBL_Model}, let $\bm{v}_P$ and $\bm{v}_T$ represent the primary and target rigid bodies, respectively, such that given the true translation vector $\bm{t}$ and rotation matrix $\bm{Q}$ relating the location and orientation of the target body relative to the primary, we have
\begin{equation}
\label{eq:SemanticRBL}
\bm{v}_T = \bm{Q}\cdot\bm{v}_P\raisebox{-2pt}{$\big|_{\bm{t}}$}\equiv \bm{Q}\cdot\bm{v}_P + \bm{t}.
\end{equation}

As illustrated in Figure \ref{fig:Sem_RBL_Model}, the entity described in equation \eqref{eq:SemanticRBL} is the image of the target, constructed in terms of a translation and rotation of the primary body to the location and orientation of the target.
In turn, the equivalence in equation \eqref{eq:SemanticRBL} refers to the fact that the end-point of the image $\bm{Q}\cdot\bm{v}_P\raisebox{-3pt}{$\big|_{\bm{t}}$}$ is at the location indicated by the vector $\bm{Q}\cdot\bm{v}_P + \bm{t}$.

It follows from the above that in presence of estimates $\hat{\bm{t}}$ and $\hat{\bm{Q}}$, we have
\begin{equation}
\label{ee:SemanticRBL_est}
\hat{\bm{v}}_T = \hat{\bm{Q}}\cdot\bm{v}_P\raisebox{-2pt}{$\big|_{\hat{\bm{t}}}$}\equiv \hat{\bm{Q}}\cdot\bm{v}_P + \hat{\bm{t}},
\end{equation}
such that the egoistic pose estimation error for a given $k$-th Monte-Carlo realization can be defined as
\begin{equation}
\label{ee:SemanticRBL_est_error}
\vartheta_k \triangleq |\hat{\bm{v}}_T^{(k)} - \bm{v}_T|_2,
\end{equation}
which yields a corresponding \ac{RMSE} (over multiple realizations) given by

\begin{equation}
\vartheta = \frac{1}{\sqrt{K}}\bigg(\sum_{k=1}^{K}|\hat{\boldsymbol{v}}_T^{(k)}-{\boldsymbol{v}_T}|_2^2\bigg)^{\!\!\frac{1}{2}}.
\end{equation}
\vspace{-2ex}

To the best of our knowledge, there is no alternative egoistic method for rigid body orientation estimation against which to compare our proposed technique.
In particular, the \ac{SotA} technique from \cite{Chen_2015} enables the joint estimation of location and orientation, but only under the assumption that full conformation matrix information is available; and introducing conformation matrix knowledge to the proposed orientation estimation method Algorithm \ref{alg:alg3} reduces it to the \ac{SotA} technique of \cite{PizzoICASSP2016}, as described in Subsection \ref{sec:Q_est}.

In view of the above, we first compare combinations of the \ac{GA} variation of Algorithm \ref{alg:alg3} ($i.e.$, the method in \cite{PizzoICASSP2016}) with either Algorithms \ref{alg:alg1} or \ref{alg:alg2}, against the \ac{SotA} alternative \cite{Chen_2015}, in order to obtain an initial reference of their relative performances, and subsequently compare the distinct egoistic methods here proposed among themselves.

\begin{figure}[H]
\centering
\includegraphics[width=\columnwidth]{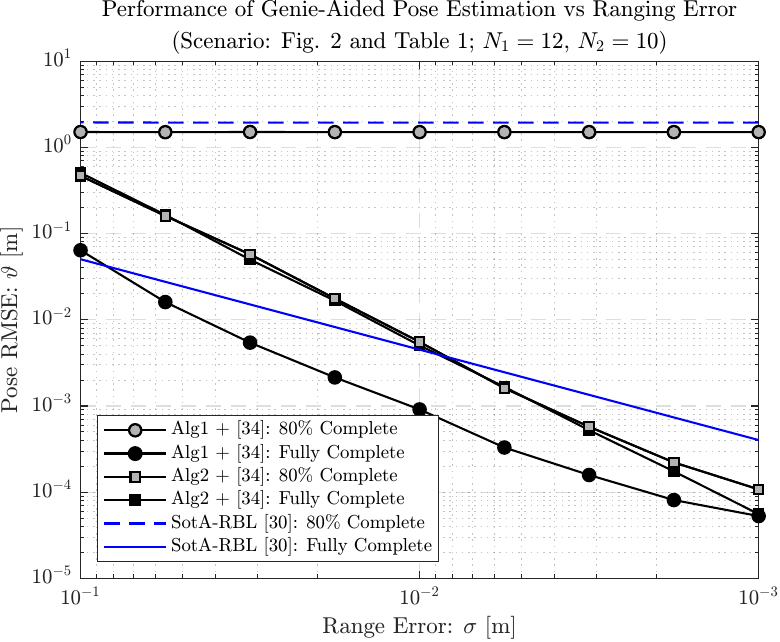}
\vspace{-5ex}
\caption{\ac{RMSE} of the genie-aided orientation estimate of the proposed methods in combination with the \ac{SotA}, compared to the pure \ac{SotA} solution for different levels of available information, over the range error $\sigma$.}
\label{fig:RMSE_rot_GA}
\end{figure}

The first set of results is offered in Figure \ref{fig:RMSE_rot_GA}, which shows the aforementioned comparisons in scenarios with fully- and 80\%-complete information.
\vspace{-2ex}

\begin{figure}[H]
    \centering
\includegraphics[width=\columnwidth]{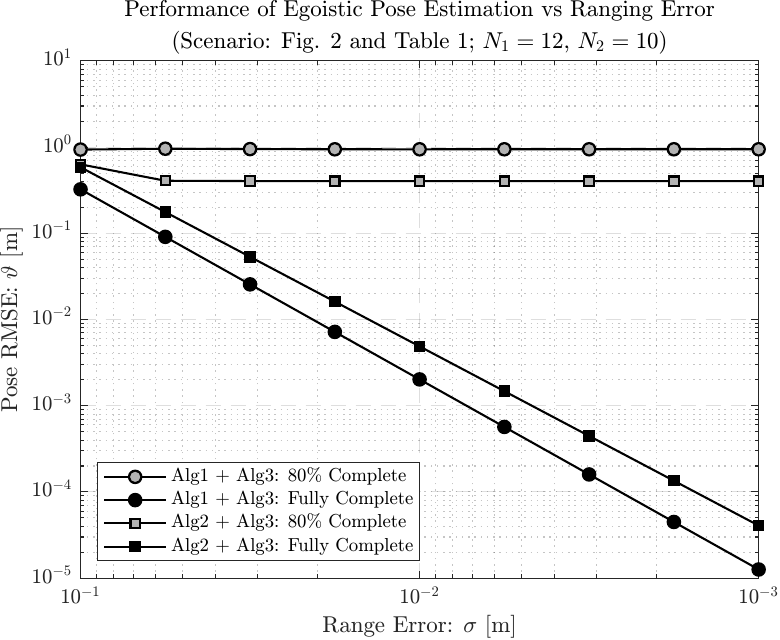}
\vspace{-4.5ex}
    \caption{\ac{RMSE} of the egoistic orientation estimate of the combinations of the proposed methods for different levels of available information, over the range error $\sigma$.}
    \label{fig:RMSE_rot_Ego}
\end{figure}

It can be observed that the results are similar to the pure translation vector estimation case evaluated in Figures \ref{fig:RMSE_MC_Conf} and  \ref{fig:RMSE_MC_J}, with Algorithm \ref{alg:alg1} performing best for the fully complete scenario and Algorithm \ref{alg:alg2} achieving the best results for $80\%$ available information.
These results are further corroborated by Figure \ref{fig:RMSE_rot_Ego}, which compares the performances of the egoistic methods, where Algorithm \ref{alg:alg1} and Algorithm \ref{alg:alg2} are combined with the rotation estimation of Algorithm \ref{alg:alg3}.
It is found that while for the fully complete scenario, the algorithms perform well, for the case of $80\%$ available information both combinations result in an error floor, with Algorithm \ref{alg:alg2} in combination with Algorithm \ref{alg:alg3} performing slightly better than Algorithm \ref{alg:alg2} in combination with Algorithm \ref{alg:alg3}.

These results indicate that matrix completion alone is not sufficient to mitigate the impact of incomplete information on the estimation of orientation matrices, such that more work is required to address that problem.
Such work is currently under pursuit and will be addressed in a follow-up article.
{Nevertheless, the results presented in this section confirm that the combination of the proposed egoistic translation and rotation estimation methods can effectively estimate the pose of a target rigid body, which is a significant step towards self-reliant rigid body localization in practical applications.}

\vspace{-2.5ex}

\subsection{Complexity Analysis}

Finally, for the sake of completeness we assess the proposed \ac{RBL} pose estimation algorithms in terms their computational complexities,  compared to \ac{SotA} methods.
In particular, we show in Table \ref{table:1} the complexity orders, in Big-O notation, of each of the compared approaches, highlighting the distinction of results under the \acf{GA} and egoistic scenarios, respectively.

We clarify that the expressions capture the complexities of all steps required prior to enabling the corresponding scenario, additionally to the solution of the optimization problem.
If matrix completion is applied to the methods, an additional complexity of order $O((N_1+N_2)^3)$ must be considered.

\begin{figure*}[t!]
\begin{minipage}{0.95\textwidth}
\begin{table}[H]
\renewcommand{\arraystretch}{1.2}
\centering
\caption{Computational Complexity of Rigid Body Pose Estimators}
\vspace{-1ex}
\begin{threeparttable}
\begin{tabular}[H]{|c|c|c|}
\hline
\raisebox{-5pt}{\textbf{Method}} & \multicolumn{2}{c|}{\textbf{Complexity Order}} \\ \cline{2-3}
                    & \textbf{Genie Aided} & \textbf{Egoistic} \\[0.5ex]
\hline\hline
Proposed Ego-RBL (Alg. 1) & $O((N_1+N_2)^3+\eta^3)$ & $O(2(N_1+N_2)^3+N_1^3)$\\[0.5ex]
\hline
Proposed Robust Ego-RBL (Alg. 2) & $O((N_1+N_2)^3+\eta^3)$ & $O(2(N_1+N_2)^3+N_1^3)$\\[0.5ex]
\hline
Proposed Rotation Estimator (Alg. 3) & $O(2\eta^3)$ (SotA \cite{PizzoICASSP2016})& $O((N_1+N_2)^3+N_1^3+2\eta^3)$\\[0.5ex]
\hline
\ac{SotA} Stationary Parameter Est. \cite{Chen_2015} & $O(\eta^3+N_1N_2)$& --- \\[0.5ex]
\hline
\end{tabular}
\label{table:1}
\begin{tablenotes}
\item[] $\eta$ denotes the dimension of the space (typically $\eta = 3$).
\item[] $N_1$ and $N_2$ denote the number of landmark points of primary and target rigid bodies, respectively.
\end{tablenotes}
\end{threeparttable}
\end{table}
\hspace{-2ex}\hrule
\end{minipage}
\end{figure*}

The results indicate that, in comparison to the non-egoistic method of \cite{Chen_2015}, which has a complexity order that is essentially quadratic on the number of landmark points defining the rigid bodies, the proposed egoistic methods have a slightly larger (cubic) complexity, which therefore can be seen as the price paid for self-reliance.
We point out, however, that since matrix completion methods also have cubic complexity, such a ``penalty'' is already imposed by practical conditions, since it is virtually impossible to ensure that complete information is acquired in real-life applications.
A preliminary contribution towards reducing the complexity of pose estimation schemes for rigid bodies can be found in \cite{vizitivWCNC2025}, and further work is also under pursuit \cite{Nic_6D_RBL}.

\section{Conclusion}

We proposed novel anchorless \ac{RBL} algorithms suitable for application in \ac{AD}, which enables a rigid body to egoistically detect the relative translation (effective distance) and orientation (relative rotation) of another rigid body, based only on measurements of the distances between sensors located at the landmark points defining the two objects, and without knowledge of the shape of the target.
Besides the self-reliance (or egoistic) feature, a key point of the proposed methods is that rotation and translation estimation can be performed independent of each other.
While the first proposed translation estimator performs well in a fully connected scenario where all distance measurements are available, the second estimator offers improved robustness to incomplete observation, even when both are furbished with matrix completion schemes.
Simulation results confirm that the proposed methods outperform \ac{SotA} alternatives.
{
While the proposed method demonstrates promising performance, several open challenges remain.
In particular, the impact of incomplete distance measurements on rotation estimation requires further investigation, as highlighted by the observed error floors in the simulations.
Future research will therefore focus on the development of more robust egoistic estimation techniques, which should also aim at reducing the computational complexity of the proposed methods to facilitate real-time applications in \ac{AD}.
Finally, rigid body tracking in dynamic scenarios also constitutes an important research direction that is of interest.
}

\appendices
\section{Corrected Relative Translation Estimator}
\label{Corrected_Est}


As mentioned in Section \ref{sec:Q_est}, the scheme proposed in \cite{PizzoICASSP2016} contains an error, which is demonstrated in this Appendix for the sake of {\color{black}completeness}.
We emphasize that, in what follows, all conditions and assumptions are the same as those in the original work, namely, that the conformation matrices $\bm{C}_1$ and $\bm{C}_2$ of both bodies are known to the estimator, and that both rigid bodies have the same size $N$.

Taking advantage of the latter, we therefore drop the subscript $_N$ from the notation of the $N \times 1$ all-ones column vector, which therefore shall be denoted here simply as $\bm{1}$.
Let us start with the Frobenius norm of the squared cross-body measurement distance matrix, which is given by
\begin{equation}
||\bm{D}_{12}||_F^2=\bm{1}^\intercal\bm{D}_{12}^{\odot2}\bm{1}.
\end{equation}

Substituting equations \eqref{eq:basic_model_one_body} and \eqref{eq:meas_full} in the above yields
\begin{subequations}
\begin{align}
||\bm{D}_{12}||_F^2 & =\bm{1}^\intercal(\boldsymbol{\psi}_1\bm{1}^{\intercal}\!+\!\bm{1}\boldsymbol{\psi}_2^{\intercal}\!-\!2\boldsymbol{S}_{1}^{\intercal}\boldsymbol{S}_{2})\bm{1}\\
&=\!N\!\sum^N_{n=1}(||\boldsymbol{s}_{1,n}||^2_2\!+\!||\boldsymbol{s}_{2,n}||^2_2)-2N\bm{1}^\intercal\times\nonumber\\
&(\bm{1}\boldsymbol{t}_{1}^\intercal\boldsymbol{t}_{2}\bm{1}^{\intercal}\!\!+\!\boldsymbol{C}_{1}^\intercal\boldsymbol{Q}_{1}^\intercal\boldsymbol{Q}_{2}\boldsymbol{C}_{2}\!+\!\boldsymbol{C}_{1}^\intercal\boldsymbol{Q}_{1}^\intercal\boldsymbol{t}_{2}\bm{1}^{\intercal}\!\!+\!\bm{1}\boldsymbol{t}_{1}^\intercal \boldsymbol{Q}_{2}\boldsymbol{C}_{2})\bm{1}.\nonumber
\end{align}

Next, using $\hat{t}=||\boldsymbol{t}_1-\boldsymbol{t}_2||_2^2=||\boldsymbol{t}_1||_2^2+||\boldsymbol{t}_2||_2^2-2\boldsymbol{t}_{1}^\intercal\boldsymbol{t}_{2}$ to find a term representing the relative translation $\hat{t}$ we obtain
\begin{align}
||\bm{D}_{12}||_F^2=&N\sum^N_{n=1}(||\boldsymbol{s}_{1,n}||^2_2+||\boldsymbol{s}_{2,n}||^2_2)\\[-3.5ex]
&+||\boldsymbol{t}_{1}||^2_2+||\boldsymbol{t}_{2}||^2_2+N^2\overbrace{||\boldsymbol{t}_1-\boldsymbol{t}_2||_2^2}^{\hat{t}}\nonumber\\[0.5ex]
&+2\bm{1}^\intercal(\boldsymbol{C}_{1}^\intercal\boldsymbol{Q}_{1}^\intercal\boldsymbol{Q}_{2}\boldsymbol{C}_{2}\!+\!\boldsymbol{C}_{1}^\intercal\boldsymbol{Q}_{1}^\intercal\boldsymbol{t}_{2}\bm{1}^{\intercal}\!+\!\bm{1}\boldsymbol{t}_{1}^\intercal \boldsymbol{Q}_{2}\boldsymbol{C}_{2})\bm{1}.\nonumber
\end{align}
\label{eq:est_steps}
\end{subequations}

Then, by rearranging equation \eqref{eq:est_steps}, the relative translation estimate becomes
\begin{align}
\label{eq:Corrected_Est_S}
\hat{t}=&\frac{-1}{N}\!\sum^N_{n=1}(||\boldsymbol{s}_{1,n}||^2_2\!+\!||\boldsymbol{s}_{2,n}||^2_2)\!+\!||\boldsymbol{t}_{1}||^2_2\!+\!||\boldsymbol{t}_{2}||^2_2\!+\!\frac{1}{N^2}||\bm{D}_{12}||_F^2\nonumber\\
&+\!\frac{2}{N^2}\bm{1}^\intercal(\boldsymbol{C}_{1}^\intercal\boldsymbol{Q}_{1}^\intercal\boldsymbol{Q}_{2}\boldsymbol{C}_{2}\!+\!\boldsymbol{C}_{1}^\intercal\boldsymbol{Q}_{1}^\intercal\boldsymbol{t}_{2}\bm{1}^{\intercal}\!+\!\bm{1}\boldsymbol{t}_{1}^\intercal \boldsymbol{Q}_{2}\boldsymbol{C}_{2})\bm{1}.\!\!
\end{align}

Finally, the estimator can be made independent of the rigid body locations $\boldsymbol{S}_1$ and $\boldsymbol{S}_2$ by making use of the relation $||\boldsymbol{s}_{1,n}||^2_2-||\boldsymbol{t}_{1}||_2^2=||\boldsymbol{c}_{1,n}||^2_2+2\boldsymbol{c}_{1,n}^\intercal\boldsymbol{Q}_1^\intercal\boldsymbol{t}_1$, which yields
\begin{eqnarray}
\label{eq:rel_tra_est}
\hat{t}=\frac{1}{N^2}||\bm{D}_{12}||_F^2-\frac{1}{N}\sum^N_{n=1}(||\boldsymbol{c}_{1,n}||^2_2+||\boldsymbol{c}_{2,n}||^2_2)+&&\\
&&\hspace{-49ex}
\text{\raisebox{-20pt}{\rotatebox{90}{$\text{missing in}\atop \text{\cite[Sec. 3.2]{PizzoICASSP2016}}$}}}
\begin{cases}
\displaystyle
+\frac{2}{N^2}\bm{1}^\intercal(\boldsymbol{C}_{1}^\intercal\boldsymbol{Q}_{1}^\intercal\boldsymbol{Q}_{2}\boldsymbol{C}_{2}+\boldsymbol{C}_{1}^\intercal\boldsymbol{Q}_{1}^\intercal\boldsymbol{t}_{2}\bm{1}^{\intercal}+\bm{1}\boldsymbol{t}_{1}^\intercal \boldsymbol{Q}_{2}\boldsymbol{C}_{2})\bm{1}\\
\displaystyle
-\frac{1}{N}\sum_{i=1}^N(2\boldsymbol{c}_{i,1}^\intercal\boldsymbol{Q}_1^\intercal\boldsymbol{t}_1+2\boldsymbol{c}_{i,2}^\intercal\boldsymbol{Q}_2^\intercal\boldsymbol{t}_2),\nonumber
\end{cases}
\end{eqnarray}
where we have highlighted the terms missing in \cite[Sec. 3.2]{PizzoICASSP2016}.

Having obtained the corrected estimator and applying the corresponding assumptions from this work, it can be observed that the method is not usable, neither in the proposed framework, not in the \ac{SotA}, since $\bm{Q_1}$, $\bm{Q_2}$, $\bm{t}_1$, and $\bm{t}_2$ are all unknowns, on which the missing terms {\color{black}depend}.



\begin{thebibliography}{10}
    \providecommand{\url}[1]{#1}
    \csname url@samestyle\endcsname
    \providecommand{\newblock}{\relax}
    \providecommand{\bibinfo}[2]{#2}
    \providecommand{\BIBentrySTDinterwordspacing}{\spaceskip=0pt\relax}
    \providecommand{\BIBentryALTinterwordstretchfactor}{4}
    \providecommand{\BIBentryALTinterwordspacing}{\spaceskip=\fontdimen2\font plus
    \BIBentryALTinterwordstretchfactor\fontdimen3\font minus
      \fontdimen4\font\relax}
    \providecommand{\BIBforeignlanguage}[2]{{%
    \expandafter\ifx\csname l@#1\endcsname\relax
    \typeout{** WARNING: IEEEtran.bst: No hyphenation pattern has been}%
    \typeout{** loaded for the language `#1'. Using the pattern for}%
    \typeout{** the default language instead.}%
    \else
    \language=\csname l@#1\endcsname
    \fi
    #2}}
    \providecommand{\BIBdecl}{\relax}
    \BIBdecl
    
    \bibitem{Nic_RBL}
    N.~Führling, H.~S. Rou, G.~T.~F. de~Abreu, D. Gonzáles ~G., and O.~Gonsa, ``Egoistic
      {MDS}-based Rigid Body Localization,'' to appear at \emph{IEEE Wireless Communications
      and Networking Conference (WCNC)}, 2025.
    
    \bibitem{burghal_2020}
    D.~Burghal \emph{et. al}, ``A
      Comprehensive Survey of Machine Learning Based Localization with Wireless
      Signals,'' 2020.
    
    \bibitem{obeidat2021review}
    H.~Obeidat, W.~Shuaieb, O.~Obeidat, and R.~Abd-Alhameed, ``A Review of Indoor
      Localization Techniques and Wireless Technologies,'' \emph{Wireless Personal
      Communications}, vol. 119, pp. 289--327, 2021.
    
    \bibitem{ShanAccess2023}
    X.~Shan, A.~Cabani, and H.~Chafouk, ``A Survey of Vehicle Localization:
      Performance Analysis and Challenges,'' \emph{IEEE Access}, vol.~11, pp.
      107\,085--107\,107, 2023.
    
    \bibitem{WangJSAC2022}
    Z.~Wang \emph{et. al}, ``Location Awareness in
      Beyond 5G Networks via Reconfigurable Intelligent Surfaces,'' \emph{IEEE
      Journal on Selected Areas in Communications}, vol.~40, no.~7, pp. 2011--2025,
      2022.
    
    \bibitem{02:00074}
    {ITU-R, International Telecommunication Union - Radiocommunication Sector},
      ``{M.2160-0: Framework and Overall Objectives of the Future Development of
      IMT for 2030 and Beyond},'' Nov. 2023.
    
    \bibitem{WangTIV2020}
    Z.~Wang, J.~Fang, X.~Dai, H.~Zhang, and L.~Vlacic, ``Intelligent Vehicle
      Self-Localization Based on Double-Layer Features and Multilayer LIDAR,''
      \emph{IEEE Transactions on Intelligent Vehicles}, vol.~5, no.~4, pp.
      616--625, 2020.
    
    \bibitem{ManickamAcces2023}
    S.~Manickam, L.~Yarlagadda, S.~P. Gopalan, and C.~L. Chowdhary, ``Unlocking the
      Potential of Digital Twins: A Comprehensive Review of Concepts, Frameworks,
      and Industrial Applications,'' \emph{IEEE Access}, vol.~11, pp.
      135\,147--135\,158, 2023.
    
    \bibitem{VoCST2016}
    Q.~D. Vo and P.~De, ``A Survey of Fingerprint-Based Outdoor Localization,''
      \emph{IEEE Communications Surveys \& Tutorials}, vol.~18, no.~1, pp.
      491--506, 2016.
    
    \bibitem{Nic:RSSI}
    N.~Führling, H.~S. Rou, G.~T.~F. de~Abreu, D.~G. G., and O.~Gonsa, ``Robust
      Received Signal Strength Indicator {(RSSI)}-Based Multitarget Localization
      via Gaussian Process Regression,'' \emph{IEEE Journal of Indoor and Seamless
      Positioning and Navigation}, vol.~1, pp. 104--114, 2023.
    
    \bibitem{Al-SadoonTAP2020}
    M.~A.~G. Al-Sadoon, R.~Asif, Y.~I.~A. Al-Yasir, R.~A. Abd-Alhameed, and P.~S.
      Excell, ``{AOA} Localization for Vehicle-Tracking Systems Using a Dual-Band
      Sensor Array,'' \emph{IEEE Transactions on Antennas and Propagation},
      vol.~68, no.~8, pp. 6330--6345, 2020.
    
    \bibitem{ZengTSP2022}
    G.~Zeng, B.~Mu, J.~Chen, Z.~Shi, and J.~Wu, ``Global and Asymptotically
      Efficient Localization From Range Measurements,'' \emph{IEEE Transactions on
      Signal Processing}, vol.~70, pp. 5041--5057, 2022.
    
    \bibitem{Yassin_2016}
    A.~Yassin \emph{et. al}, ``Recent Advances in Indoor Localization: A Survey on
      Theoretical Approaches and Applications,'' \emph{IEEE Communications Surveys
      \& Tutorials}, vol.~19, no.~2, pp. 1327--1346, 2017.
    
    \bibitem{Zhang_2021}
    J.~A. Zhang, F.~Liu, C.~Masouros, R.~W. Heath, Z.~Feng, L.~Zheng, and
      A.~Petropulu, ``An Overview of Signal Processing Techniques for Joint
      Communication and Radar Sensing,'' \emph{IEEE Journal of Selected Topics in
      Signal Processing}, vol.~15, no.~6, pp. 1295--1315, 2021.
    
    \bibitem{Rayan_2024}
    K.~R.~R. Ranasinghe, H.~S. Rou, and G.~T.~F. de~Abreu, ``Fast and Efficient
      Sequential Radar Parameter Estimation in {MIMO-OTFS} Systems,'' in \emph{IEEE
      International Conference on Acoustics, Speech and Signal Processing
      (ICASSP)}, 2024.
    
    \bibitem{Rayan_Journal}
    K.~R.~R. Ranasinghe, H.~S. Rou, G.~T.~F. de~Abreu, T.~Takahashi, and K.~Ito,
    ``Joint Channel, Data and Radar Parameter Estimation for AFDM Systems in Doubly-Dispersive Channels,'' \textit{IEEE Transactions on Wireless Communications}, vol.~--, no.~--, pp.~1--1, 2024,
    
    \bibitem{WangTSP2020}
    Y.~Wang, G.~Wang, S.~Chen, K.~C. Ho, and L.~Huang, ``An Investigation and
      Solution of Angle Based Rigid Body Localization,'' \emph{IEEE Transactions on
      Signal Processing}, vol.~68, pp. 5457--5472, 2020.
    
    \bibitem{Nic_RBL_WP}
    N.~Führling, H.~S. Rou, G.~T.~F. de~Abreu, D. Gonzáles G., and O.~Gonsa,
      ``Soft-connected Rigid Body Localization: State-of-the-art and Research
      Directions for {6G},'' \emph{arXiv preprint arXiv:2309.05002}, 2023.
    
    \bibitem{FuehrlingV2X2024}
    \BIBentryALTinterwordspacing
    N.~F{\"u}hrling, H.~S. Rou, G.~T.~F. de~Abreu, D. Gonzáles G., and O.~Gonsa,
      ``Enabling Next-Generation {V2X} Perception: Wireless Rigid Body Localization
      and Tracking,'' \emph{arXiv preprint arXiv:2408.00349}, 2024.
    \BIBentrySTDinterwordspacing
    
    \bibitem{AHMED_2020}
    \BIBentryALTinterwordspacing
    F.~Ahmed, M.~Phillips, S.~Phillips, and K.-Y. Kim, ``Comparative Study of
      Seamless Asset Location and Tracking Technologies,'' \emph{Procedia
      Manufacturing}, 30th International Conference
      on Flexible Automation and Intelligent Manufacturing, vol.~51, pp. 1138--1145, 2020:
    \BIBentrySTDinterwordspacing
    
    \bibitem{Bruk_2023}
    B.~Gebregziabher, ``Multi Object Tracking for Predictive Collision Avoidance,'' \emph{arXiv preprint arXiv:2307.02161},
      2023.
    
    \bibitem{eckenhoff_2019}
    K.~Eckenhoff, Y.~Yang, P.~Geneva, and G.~Huang, ``Tightly-Coupled
      Visual-Inertial Localization and {3-D} Rigid-Body Target Tracking,''
      \emph{IEEE Robotics and Automation Letters}, vol.~4, no.~2, pp. 1541--1548,
      2019.
    
    \bibitem{Huang_2022}
    Y.~Huang, J.~Du, Z.~Yang, Z.~Zhou, L.~Zhang, and H.~Chen, ``A Survey on
      Trajectory-Prediction Methods for Autonomous Driving,'' \emph{IEEE
      Transactions on Intelligent Vehicles}, vol.~7, no.~3, pp. 652--674, 2022.
    
    \bibitem{Huang_2019}
    J.~Huang, S.~Yang, Z.~Zhao, Y.-K. Lai, and S.~Hu, ``{ClusterSLAM}: A {SLAM}
      Backend for Simultaneous Rigid Body Clustering and Motion Estimation,'' in
      \emph{IEEE/CVF International Conference on Computer Vision (ICCV)},
      pp. 5874--5883, 2019.
    
    \bibitem{Barros_2022}
    \BIBentryALTinterwordspacing
    A.~Macario~Barros, M.~Michel, Y.~Moline, G.~Corre, and F.~Carrel, ``A
      Comprehensive Survey of Visual {SLAM} Algorithms,'' \emph{Robotics}, vol.~11,
      no.~1, 2022.
    \BIBentrySTDinterwordspacing
    
    \bibitem{Bavle_2023}
    \BIBentryALTinterwordspacing
    H.~Bavle, J.~L. Sanchez-Lopez, C.~Cimarelli, A.~Tourani, and H.~Voos, ``From
      {SLAM} to Situational Awareness: Challenges and Survey,'' \emph{Sensors},
      vol.~23, no.~10, 2023.
    \BIBentrySTDinterwordspacing

    {
    \bibitem{Yin2024}
    H. Yin, X. Xu, S. Lu, X. Chen, R. Xiong, S. Shen, C. Stachniss, and Y. Wang, 
    ``A Survey on Global LiDAR Localization: Challenges, Advances and Open Problems,'' 
    \emph{International Journal of Computer Vision}, vol.~132, no.~8, pp.~3139--3171, Aug. 2024.


    \bibitem{chen2021lidarlocalization}
X. Chen, I. Vizzo, T. Läbe, J. Behley, and C. Stachniss, 
``Range Image-Based LiDAR Localization for Autonomous Vehicles,'' 
in \emph{Proc. IEEE Int. Conf. on Robotics and Automation (ICRA)}, 
Xi'an, China, pp.~5802--5808, 2021.



\bibitem{Elhousni_2020}
M. Elhousni and X. Huang, 
``A Survey on 3D LiDAR Localization for Autonomous Vehicles,'' 
in \emph{Proc. IEEE Intelligent Vehicles Symp. (IV)}, pp.~1879--1884, 2020.

    }
    
    \bibitem{Chen_2015}
    S.~Chen and K.~C. Ho, ``Accurate Localization of a Rigid Body Using Multiple
      Sensors and Landmarks,'' \emph{IEEE Transactions on Signal Processing},
      vol.~63, no.~24, pp. 6459--6472, 2015.
    
{\color{black}    \bibitem{9447218}
G. Wang, K. C. Ho, and X. Chen, 
``Bias Reduced Semidefinite Relaxation Method for 3-D Rigid Body Localization Using AOA,'' 
\emph{IEEE Transactions on Signal Processing}, vol.~69, pp.~3415--3430, 2021.

\bibitem{9904904}
X. Wu, Q. Lin, and H. Qi, 
``Cooperative Multiple Rigid Body Localization via Semidefinite Relaxation Using Range Measurements,'' 
\emph{IEEE Transactions on Signal Processing}, vol.~70, pp.~4788--4803, 2022.
}

    \bibitem{Bras_2016}
    S.~Brás, M.~Izadi, C.~Silvestre, A.~Sanyal, and P.~Oliveira, ``Nonlinear
      Observer for {3D} Rigid Body Motion Estimation Using Doppler Measurements,''
      \emph{IEEE Transactions on Automatic Control}, vol.~61, no.~11, pp.
      3580--3585, 2016.
    
    \bibitem{PizzoICASSP2016}
    A.~Pizzo, S.~P. Chepuri, and G.~Leus, ``Towards Multi-Rigid Body
      Localization,'' in \emph{IEEE International Conference on Acoustics, Speech
      and Signal Processing (ICASSP)}, pp. 3166--3170, 2016.
    
    \bibitem{Chepuri_2013}
    S.~P. Chepuri, A.~Simonetto, G.~Leus, and A.-J. van~der Veen, ``Tracking
      Position and Orientation of a Mobile Rigid Body,'' in \emph{5th IEEE
      International Workshop on Computational Advances in Multi-Sensor Adaptive
      Processing (CAMSAP)}, pp. 37--40, 2013.

     {\color{black} \bibitem{9186663}
Y. Wang, G. Wang, S. Chen, K. C. Ho, and L. Huang, 
``An Investigation and Solution of Angle Based Rigid Body Localization,'' 
\emph{IEEE Transactions on Signal Processing}, vol.~68, pp.~5457--5472, 2020.
}

    \bibitem{vizitivWCNC2025}
    \BIBentryALTinterwordspacing
    V.~Vizitiv, H.~S. Rou, N.~Führling, and G.~T.~F. de~Abreu, ``Belief
      Propagation-based Rotation and Translation Estimation for Rigid Body
      Localization,'' to appear at \emph{IEEE Wireless Communications
      and Networking Conference (WCNC)}, 2025. Available at \emph{arXiv:2407.09232}, 2024. 
    \BIBentrySTDinterwordspacing
    
    \bibitem{Torgerson_1952}
    \BIBentryALTinterwordspacing
    W.~S. Torgerson, ``Multidimensional Scaling: I. Theory and Method,''
      \emph{Psychometrika}, vol.~17, no.~4, pp. 401--419, Dec. 1952.
    \BIBentrySTDinterwordspacing
    
    \bibitem{OPP_1966}
    P.~Schönemann, ``A Generalized Solution of the Orthogonal Procrustes
      Problem,'' \emph{Psychometrika}, vol.~31, no.~1, pp. 1--10, 1966.
    
    \bibitem{Fang2012}
    \BIBentryALTinterwordspacing
    H.~Fang and D.~P. O'Leary, ``Euclidean Distance Matrix Completion Problems,''
      \emph{Optimization Methods and Software}, vol.~27, no. 4-5, pp. 695--717,
      2012. 
    \BIBentrySTDinterwordspacing
    
    \bibitem{Nguyen2019}
    L.~T. Nguyen, J.~Kim, and B.~Shim, ``Low-Rank Matrix Completion: A Contemporary
      Survey,'' \emph{IEEE Access}, vol.~7, pp. 94\,215--94\,237, 2019.
    
    \bibitem{Fan2024}
    Y.~Fan and M.~Pesavento, ``Localization in Sensor Networks Using Distributed
      Low-Rank Matrix Completion,'' in \emph{IEEE International Conference on
      Acoustics, Speech and Signal Processing (ICASSP)}, pp.
      12\,861--12\,865, 2024.
    
    \bibitem{Williams_2001}
    C.~K.~I. Williams and M.~Seeger, ``Using the {Nystr\"{o}m} Method to Speed Up
      Kernel Machines,'' in \emph{Proceedings of the 13th Int. Conf.
      on Neural Information Processing Systems}, ser. NIPS'00.\hskip 1em plus 0.5em
      minus 0.4em\relax Cambridge, MA, USA: MIT Press, p. 661–667, 2000.
    
  {
      \bibitem{Boyd_Vandenberghe_2004}
S. Boyd and L. Vandenberghe, 
\emph{Convex Optimization}. 
Cambridge, U.K.: Cambridge University Press, 2004.
}

    \bibitem{Nocedal1999}
    J.~Nocedal and S.~J. Wright, \emph{Numerical Optimization}, ser. Springer
      Series in Operations Research and Financial Engineering.\hskip 1em plus 0.5em
      minus 0.4em\relax Springer New York, NY, 1999.
    
    \bibitem{Ruder2016}
    S.~Ruder, ``An Overview of Gradient Descent Optimization Algorithms,''
      \emph{arXiv preprint arXiv:1609.04747}, 2016.
    
    \bibitem{cvx}
    M.~Grant and S.~Boyd, ``{CVX}: Matlab Software for Disciplined Convex
      Programming, Version 2.1,'' \url{https://cvxr.com/cvx}, Mar. 2014.
    
    \bibitem{gb08}
    ------, ``Graph Implementations for Nonsmooth Convex Programs,'' in
      \emph{Recent Advances in Learning and Control}, ser. Lecture Notes in Control
      and Information Sciences, V.~Blondel, S.~Boyd, and H.~Kimura, Eds.\hskip 1em
      plus 0.5em minus 0.4em\relax Springer-Verlag Limited, pp. 95--110, 2008.
    
    \bibitem{MalekianSenJ2018}
    R.~Malekian \emph{et al.}, ``Guest Editorial
      Special Issue on Sensor Technologies for Connected Cars: Devices, Systems and
      Modeling,'' \emph{IEEE Sensors Journal}, vol.~18, no.~12, pp. 4775--4776,
      2018.
    
    \bibitem{RankEDM}
    W.~Glunt, T.~L. Hayden, S.~Hong, and J.~Wells, ``An Alternating Projection
      Algorithm for Computing the Nearest Euclidean Distance Matrix,'' \emph{SIAM
      Journal on Matrix Analysis and Applications}, vol.~11, no.~4, pp. 589--600,
      1990.
    
    \bibitem{OptSpace}
    R.~H. Keshavan, A.~Montanari, and S.~Oh, ``Matrix Completion from a Few
      Entries,'' \emph{IEEE Transactions on Information Theory}, vol.~56, no.~6,
      pp. 2980--2998, 2010.
    
    \bibitem{Mazumder_2010}
    R.~Mazumder, T.~Hastie, and R.~Tibshirani, ``Spectral Regularization Algorithms
      for Learning Large Incomplete Matrices,'' \emph{Journal of Machine Learning
      Research}, vol.~11, no.~80, pp. 2287--2322, 2010.
    
    \bibitem{Yao_2019}
    Q.~Yao and J.~T. Kwok, ``Accelerated and Inexact Soft-Impute for Large-Scale
      Matrix and Tensor Completion,'' \emph{IEEE Transactions on Knowledge and Data
      Engineering}, vol.~31, no.~9, pp. 1665--1679, 2019.
    
      {\bibitem{fuehrling2025fundamentallimits}
N. F\"uhrling, I. A. Morales Sandoval, G. T. F. de Abreu, G. Seco-Granados, D. González G., and O. Gonsa,
``Fundamental Limits of Rigid Body Localization,''
\emph{arXiv preprint arXiv:2507.22573}, 2025.}

    \bibitem{Macedo_2013}
    H.~D. {Macedo} and J.~N. {Oliveira}, ``{Typing Linear Algebra: A
      Biproduct-Oriented Approach},'' \emph{arXiv preprint arXiv:1312.4818},
      Dec. 2013.
    
    \bibitem{jolliffe2002principal}
    I.~T. Jolliffe, \emph{Principal Component Analysis for Special Types of
      Data}.\hskip 1em plus 0.5em minus 0.4em\relax Springer, 2002.
    
    \bibitem{hastie2009elements}
    T.~Hastie, R.~Tibshirani, J.~H. Friedman, and J.~H. Friedman, \emph{The
      Elements of Statistical Learning: Data Mining, Inference, and
      Prediction}.\hskip 1em plus 0.5em minus 0.4em\relax Springer, vol.~2, 2009.
    

    \bibitem{Nic_6D_RBL}
      N. Führling, V. Vizitiv, K. R. R. Ranasinghe, H. S. Rou, G. T. F. de Abreu, D. González G., and O. Gonsa, 
      ``6D Rigid Body Localization and Velocity Estimation via Gaussian Belief Propagation,'' 
      \emph{arXiv preprint arXiv:2412.12133}, 2024.



    

    \end{thebibliography}


\begin{IEEEbiography}[{\includegraphics[width=1in,height=1.25in,clip,keepaspectratio]{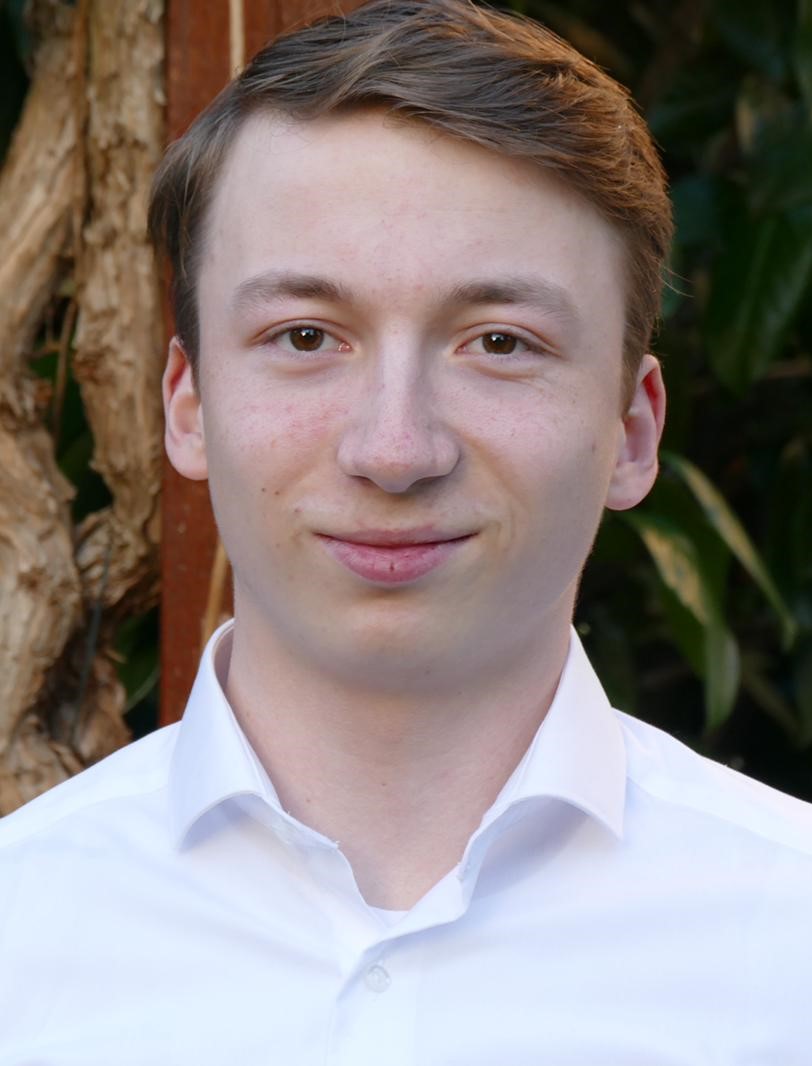}}]{Niclas F{\"u}hrling}
(Graduate Student Member, IEEE) received the B.Sc. degree in electrical and computer engineering from Jacobs University Bremen, Bremen, Germany in 2022 and the M.Sc. degree in electrical engineering with the University of Bremen in 2024, with a focus on communication and information technology. He is currently pursuing the Ph.D. degree in Electrical Engineering with Constructor University Bremen, Germany, while working on a research project  focusing on 6G connectivity. His current research interests are wireless communications, signal processing and wireless localization.
\end{IEEEbiography}
\begin{IEEEbiography}[{\includegraphics[width=1in,height=1.25in,clip,keepaspectratio]{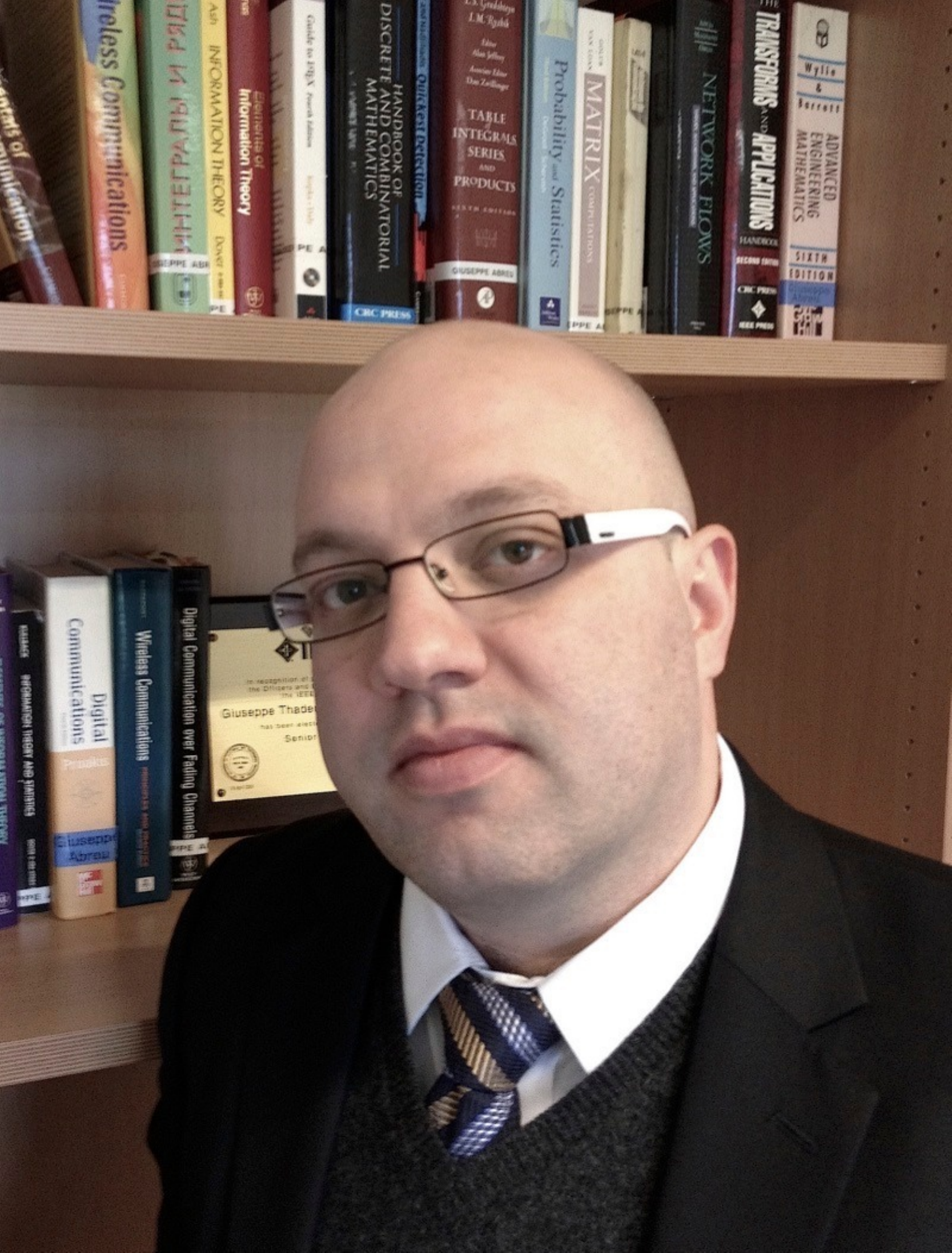}}]{Giuseppe Thadeu Freitas de Abreu} (Senior Member, IEEE) received the B.Eng. degree in electrical engineering and the specialization Latu Sensu
degree in telecommunications engineering from the
Universidade Federal da Bahia (UFBa), Salvador,
Bahia, Brazil in 1996 and 1997, respectively, and
the M.Eng. and D.Eng. degrees in physics, electrical,
and computer engineering from Yokohama National
University, Japan, in March 2001 and March 2004,
respectively. He was a postdoctoral fellow and later
an adjunct professor (docent) in statistical signal
processing and communications theory at the Department of Electrical and
Information Engineering, University of Oulu, Finland from 2004 to 2006 and
from 2006 to 2011, respectively. Since 2011, he has been a professor of
electrical engineering at Constructor University (formerly known as Jacobs
University), Bremen, Germany. From April 2015 to August 2018, he also
simultaneously held a full professorship at the Department of Computer
and Electrical Engineering, Ritsumeikan University, Japan. His research
interests include communications and signal processing, communications
theory, estimation theory, statistical modeling, wireless localization, cognitive radio, wireless security, MIMO systems, ultrawideband and millimeter
wave communications, full-duplex and cognitive radio, compressive sensing,
energy harvesting networks, random networks, connected vehicles networks,
electro-magnetic signal processing, quantum computing for signal processing,
metasurfaces for wireless systems and many other topics. Prof. Abreu has
received various awards and prestigious fellowships, including the Uenohara
Award from Tokyo University in 2000, the prestigious JSPS, Heiwa Nakajima,
and NICT (twice) fellowships in 2010, 2013, 2015, and 2018, as well
as Best Paper award at several international conferences, and of the Best
Paper award by the Japanese Chapter of the IEEE Signal Processing Society
in 2023. He served as an associate editor for the IEEE Transactions on
Wireless Communications from 2009 to 2014 and the IEEE Transactions
on Communications from 2014 to 2017; as an executive editor for IEEE
Transactions on Wireless Communications from 2017 to 2021, and as an
editor IEEE Communications Letters from 2021 to 2023. He is currently
serving as an editor to the IEEE Signal Processing Letters and to the IEEE
Open Journal of the Communications Society.
\end{IEEEbiography}
\begin{IEEEbiography}[{\includegraphics[width=1in,height=1.25in,clip,keepaspectratio]{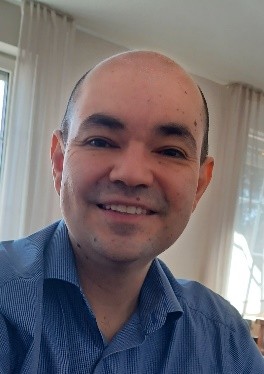}}]{David González G., PhD.}
(S'06-M'15-SM'18) has a master in Mobile Communications and PhD in Signal Theory and Communications from the Universitat Politècnica de Catalunya, Spain. He has served as post-doctoral fellow in Aalto University, Finland (2014-2017). He also served as Research Engineer with Panasonic Research and Development Center, Germany. Since 2018, David is with Continental Automotive Technologies, Germany, where he conducts and manages research projects focused on diverse aspects of vehicular communications (V2X), integrated sensing and communications (ISAC), and automotive applications for 5G-Advanced and 6G. David participates as delegate in 3GPP RAN1, 5GAA, and ETSI ISG ISAC. 
\end{IEEEbiography}
\begin{IEEEbiography}[{\includegraphics[width=1in,height=1.25in,clip,keepaspectratio]{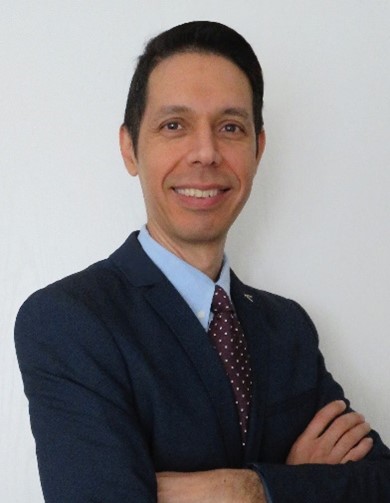}}]{Osvaldo Gonsa}
received the Ph.D. degree in electrical and computer engineering from Yokohama National University, Japan, in 1999, and the M.B.A. degree from the Kempten School of Business, Germany, in 2012. Since 1999 he has worked in research and standardization in the areas of core and radio access network. He is currently the Head of the Wireless Communications Technologies Group, Continental AG, Frankfurt, Germany. And since 2020 also serves as a member for the GSMA Advisory Board for automotive and the 6GKom Project of the German Federal Ministry of Education and Research.
\end{IEEEbiography}

\end{document}